\documentclass[a4paper,12pt]{article}

\usepackage{bbm}
\usepackage[utf8]{inputenc}
\usepackage{placeins}
\usepackage{amssymb,amsmath,amsfonts}
\usepackage[colorlinks,linkcolor=purple,citecolor=teal]{hyperref}
\usepackage{dsfont}
\usepackage{color}
\usepackage{graphicx}
\usepackage{hyphenat}
\usepackage{wrapfig}
\usepackage{empheq}
\usepackage{textcomp}
\usepackage[caption=false]{subfig}
\usepackage{rotating}
\usepackage{wrapfig}
\usepackage{pdfpages}
\usepackage{verbatim}
\usepackage{setspace}
\usepackage{array}
\usepackage{caption}
\usepackage[pdf]{pstricks}
\usepackage{upgreek}
\usepackage{cmll}
\usepackage{latexsym}
\usepackage{braket}
\usepackage{cite}
\usepackage{epsfig}
\usepackage[left=2.4cm,top=3.3cm,right=2.4cm,bottom=3.3cm,bindingoffset=0cm]{geometry}
\usepackage[titletoc,page]{appendix}
\usepackage{multicol}
\usepackage{youngtab}

\newcommand{\beq}{\begin{eqnarray}}
\newcommand{\eeq}{\end{eqnarray}}
\newcommand{\bea}{\begin{eqnarray}}
\newcommand{\eea}{\end{eqnarray}}
\newcommand{\be}{\begin{equation}}
\newcommand{\ee}{\end{equation}}

\def\brc{\langle}
\def\ckt{\rangle}

\def\de{\partial}

\def\Tr{\qopname\relax o{Tr}}

\numberwithin{equation}{section}

\numberwithin{equation}{section}

\begin{document}
 

\title{The Quantum Ratio } 


\author{    Kenichi Konishi$^{(1,2)}$, 
Hans-Thomas Elze$^{(1)}$   \\[13pt]
{\em \footnotesize
$^{(1)}$Department of Physics ``E. Fermi", University of Pisa}\\
{\em \footnotesize
Largo Pontecorvo, 3, Ed. C, 56127 Pisa, Italy}\\[2pt]
 {\em \footnotesize
$^{(2)}$INFN, Sezione di Pisa,    
Largo Pontecorvo, 3, Ed. C, 56127 Pisa, Italy}
\\[-5pt]
\\[1pt] 
{ \footnotesize  kenichi.konishi@unipi.it,  elze@df.unipi.it}  
}
\date{}

\maketitle

\begin{abstract}

The concept of {\it  quantum ratio}  emerged  in the recent efforts to
 understand how Newton's equations appear for the center of mass  (CM)   of an isolated  macroscopic body at finite body-temperatures,  as the first approximation to quantum-mechanical equations.
It  is defined as $Q\equiv   R_q/L_0$,   where  the quantum fluctuation range $R_q$ is the spatial extension of the pure-state CM  wave function,  whereas  $L_0$ stands for  the body's linear  size  (the space support of the internal, bound-state wave function).   The two cases   $R_q /L_0  \lesssim  1$ or $R_q/ L_0  \gg 1$,  roughly  correspond to  
 the body's CM behaving classically or quantum mechanically, respectively. 
 In the present note we elaborate more  on this concept, illustrating it in several examples.  
  An important notion following from introduction of the quantum ratio is that the elementary particles  (thus the electron and the photon) are quantum mechanical, even when the environment-induced decoherence turns them into a mixed state. Decoherence and classical state should not be identified. This simple observation,  further illustrated by the consideration of  a few atomic or molecular processes, 
   may have significant implications on the way quantum mechanics works in biological systems.

\end{abstract}

\vskip 30 pt


\bigskip

\newpage

\tableofcontents

\newpage

\section{Introduction:  the quantum ratio \label{QRintro} }

The concept of quantum ratio emerged during the efforts to understand the conditions  under which  the center of mass (CM) of an isolated macroscopic body
possesses a unique classical trajectory.    
It is defined as 
\be    Q \equiv   \frac{R_q}{L_o}\;,    \label{QR}
\ee
where $R_q$ is the quantum fluctuation range  of the CM of the body under consideration,  and $L_0$ is  the body's  (linear) size.  
The criterion proposed to tell whether the body behaves quantum mechanically or  classically   is     \cite{KK2} 
\be  Q    \gg 1\;, \qquad  {\rm (quantum)},
 \label{quantum}   \ee
 or 
  \be      Q   \lesssim   1    \;,\qquad  {\rm (classical)}\;,     \label{classical} 
 \ee
 respectively.

 Let us assume that the total wave function of the body  has a factorized form, 
\be   \Psi({\bf r}_1,  {\bf r}_2, \ldots  {\bf r}_N) =   \Psi_{CM}({\bf R}) \,\psi_{int}({\hat  {\bf r}}_1,  {\hat  {\bf r}}_2, \ldots   {\hat  {\bf r}}_{N-1}) \;.   \label{factorization}
\ee
where $  \Psi_{CM}$ is the CM wave function,  the $N$-body bound state is described by the internal wave function $\psi_{int}$. $\{ {\hat  {\bf r}}_1,  {\hat  {\bf r}}_2, \ldots   {\hat  {\bf r}}_{N-1}\}$  are the internal positions of the component atoms or molecules, ${\bf R}$ is the CM position and    ${\bf r}_i =  {\bf R}+  {\hat  {\bf r}}_i$  ($i=1,2,\ldots,N$).      In the case of a macroscopic body $N$ can be as large as $N\sim 10^{25}, 10^{50},$  etc.

\subsection{The size of the body  \label{size}} 
$L_0$ is determined by  $\psi_{int}$.     A possible definition of  $L_0$ is
\be    L_0  =  {\rm Max}_i \,  {\bar  r_i}\;, \qquad       {\bar  r_i} \equiv (\brc  \psi_{int} | ({\hat {\bf r}}_i)^2   | \psi_{int} \ckt )^{1/2} \;, \label{sizeL}
\ee
but the detailed definition is not important here. $L_0$  is the spatial support (extension) of the internal wave function describing the bound state  \footnote{A macroscopic body, and from some scales upwards, 
might well be described as a classical bound state, due to gravitational or electromagnetic forces, but their size is  always well defined.
}.    It is  the (linear) size of the body.    Even though $L_0$ might somewhat depend on the body temperature $T$  (the average internal excitation energy)  it is well defined 
even in the $T\to 0$  limit.  It represents the extension of the ground-state wave function of the bound state describing the body.

For an atom the definition  (\ref{sizeL}) gives  correctly  the outmost orbit in the electronic configuration. 
  $L_0$ varies  from $0.5 $ $\AA$  \,    to hundreds of $\AA$ \,   for atoms and  molecules.   The atomic nuclei (composed of protons and neutrons)   which are bound  more strongly by short-range nuclear forces,    have smaller size  $L_0$  of the order of Fermi $\sim 10^{-13}$ cm.     
 As for mesoscopic to macroscopic bodies  $L_0$ varies vastly, depending on their composition, types of the forces which bind them and their particular molecular or crystalline structures.  For the earth  (the radius) $L_0 \sim 6400$ km.

An exception is the case of the elementary particles: they have  $L_0=0$.  This has a simple implication, according to (\ref{quantum}):   having  $Q=\infty$,   {\it   the elementary particles are  quantum mechanical}. 
 
One might argue that the length scales  (i.e.,  small or large)  are relative concepts in physics:  at distance  much larger than $L_0$, any body looks pointlike.  
More generally, changing the scales of distances or energies, physics might  look similar.  A more rigorous formulation of this idea (the scale invariance) is that of the renormalization group in relativistic quantum field theories in four dimensions (e.g.,  theories of the fundamental interactions  \cite{Weinberg,Salam,Glashow,GellMann}), or in lower dimensional models of critical phenomena  \cite{Wilson2}.

Scale invariance  holds if the system possesses no fixed  length scale  \footnote{Nonrelativistic quantum
 mechanics, having only $\hbar$ with the dimension of an action as the fundamental constant in its formulation,  shares this property \cite{KK1}. The so-called quantum nonlocality is one of its consequences.  
However in specific problems, the masses and  the potential break  explicitly the scale invariance in general. For a class of the potentials, such as the delta-function or $1/r^2$ potentials in $D=2$ space dimensions,  the system possesses an exact scale invariance   \cite{Jackiw}.  }.  
      From the point of view of the theory of the fundamental interactions,   the absence of a fixed length scale means that physics at low energies does not 
     depend on the ultraviolet cutoff $\Lambda$  we need to introduce to regularize (and renormalize)  the theory  because of the ultraviolet divergences present.  In other words,   the theory is of renormalizable type,   that is,  a quantum field theory without any a priori mass (or length) parameter.

      For the questions of interest in the present  work,  however,  it is important, and we do know,  that   {\it  the world we live in   has  definite length scales, such as Bohr's radius, and the size of the atomic nuclei.}   In other words,   the terms such as microscopic (from elementary particles, nuclei, atoms,  to molecules) 
  and macroscopic (much larger than these) have a well-defined, concrete meaning. 
 
  These fixed  sizes (or length scales) characterizing our world are set by the fundamental constants  of Nature ($\hbar$, $e$, $c$) and by the parameters  in 
the   theory of the fundamental  (strong and electroweak) interactions  \cite{Weinberg,Salam,Glashow,GellMann},
 namely, the quark and lepton masses,  $W$ and $Z$ masses.  See Sec.~\ref{EP} more about this.

 \subsection{Quantum range $R_q$} 
The quantum fluctuation range $R_q$ is determined by   $\Psi_{CM}({\bf R})$.
 In principle, it is just the (spatial) extension of the pure-state wave function  $\Psi_{CM}({\bf R})$  describing the CM of the body.   
But  it  is a much more complex quantity than $L_0$:   it depends on many factors.   
     In quantum mechanics (QM)  there is no a priori upper limit to  $R_q$ \footnote{This is  another consequence of the fact that QM laws  contain no fundamental constant with the dimension of a length.  }. 
     Take for instance the wave function of a free particle,  $\psi$.   It might be thought that the normalization condition $||\psi ||=1$ necessarily sets a
finite quantum fluctuation range, but it is not so.  As is well known  (Weyl's criterion), a particle can be in a state arbitrarily close to  a plane-wave state, 
\be \psi \propto   e^{i {\bf p} \cdot {\bf r} / \hbar} \;,   \label{plane} 
\ee
i.e.,  in a momentum eigenstate, which has $R_q=\infty$.

  Given a body, $R_q$  will in general depend on the internal structures,  excitation modes and  on its body-temperature. These cause the  self-induced (or  thermal)  decoherence,     due to the emission of photons which carry away information, and seriously reduce   $R_q$.    If the body  is not isolated,   its $R_q$  is  severely affected by the environment-induced decoherence \cite{Joos1}-\!\cite{Zurek2},  on the surrounding temperature, flux, etc.  $R_q$ also depends  on the external  electromagnetic fields,  which may split the wave packets as in the Stern-Gerlach set-up,  or  on possible quantum-mechanical correlations (entanglement) among distant particles. 
   $R_q$   may depend also on time.

An important question concerns  the {\it  width of the wave packet}   of  the CM of an isolated (microscopic or macroscopic) particle, $\Delta_{CM}$.
It should not be confused with $L_0$.  
Being the spread of a single-particle wave function,   $\Delta_{CM}$  {\it is} a measure of  the quantum fluctuation range  
\be    R_q \gtrsim  \Delta_{CM} \;,     \label{together}     \ee        
but  $R_q$ can be much larger than  $\Delta_{CM}$ in general.

As for the relation between  $\Delta_{CM}$  and $L_0$, $\Delta_{CM}$ corresponds to the uncertainty of the CM position of the body.
   For a macroscopic body, an experimentalist  who is capable to measure and determine  its size $L_0$ with some precision, will certainly  be able to measure the CM position  $R$ with  
\be   \Delta_{CM}  \lesssim   L_0\;,\qquad   {\rm or\,\, even\,\, with } \qquad     \Delta_{CM}  \ll  L_0 \;. \label{welldefined}  \ee
    Nevertheless,  such a relation neither holds necessarily nor is required in general.

      A macroscopic body, especially at exceedingly low temperatures near $T=0$,  may well be   in  a state of position uncertainty (the width of the wave packet)
\be       \Delta_{CM}  \gg      L_0\;.    \label{class} 
\ee
Such a system  is seen, by using (\ref{together}),  to have   $Q \gg 1$, so is quantum mechanical.   Many attempts to realize macroscopic quantum states experimentally, bringing the system temperatures down to close to $T=0$, have been made recently  \cite{Leggett}-\cite{Aaron}.

Vice versa,  a well-defined  CM position   (\ref{welldefined}) set up  at time  $t=0$, does not in itself tell whether the system will behave 
 quantum-mechanically
or classically.

A free wave packet of  an atom or molecule with the initial position uncertainty  $\Delta_{CM}$,  will quickly diffuse (the diffusion rate depends on the mass) and will acquire 
$R_q \sim \Delta_{CM} \gg L_0.$   See Table~\ref{diffusion} taken from \cite{KK2}.    
 In the Stern-Gerlach set-up,  with an inhomogeneous magnetic field,    the (transverse) wave packet of an atom or a molecule with spin  will be spilt  in  two or more  wave packets, which can get separated even by a macroscopic distance  ($R_q$), such that  $R_q  \gg L_0$, $Q \gg 1$
 (see  Sec.~\ref{SG} for  more about it).

 \begin{table}
  \centering 
  \begin{tabular}{|c|c|c|  }
\hline
 particle   &   mass  (in $g$)  &    diffusion time   (in $s$)  \\   \hline
 electron   &   $9  \cdot  10 ^{-28}  $  &     $10^{-8}   $   \\
   hydrogen atom   &   $1.6  \cdot  10 ^{-24}  $    &    $1.6 \cdot 10^{-5}  $  \\
   $C_{70}$ fullerene  &   $8   \cdot  10 ^{-22} $  &     $8 \cdot 10^{-3}$    \\
   a stone of $1g$    &     $  1  $    &     $10^{19}      $  \\  
\hline
\end{tabular}
  \caption{ \footnotesize  Diffusion  time  of the free wave packet for different particles. Conventionally, we take the initial wave packet size of $1 \mu =10^{-6} m$, and define 
  the diffusion time  as $\Delta t$ needed for doubling its size.  For a  macroscopic particle of $1 g$, the doubling time,  $   10^{19}  {\rm sec}   \sim 10^{11}  {\rm yrs}$,
   exceeds  the age of the universe.
       }\label{diffusion}
\end{table}

On the other hand,  a macroscopic body does not diffuse (see Table~\ref{diffusion}).   Its CM  wave packet does not split under an inhomogeneous magnetic field,  either \cite{KK2}.   
Therefore,  if the CM position of a macroscopic body is measured with precision (\ref{welldefined})  at time $t=0$, the relation  $R_q \lesssim L_0$  ($Q \lesssim 1$ )   is maintained in time.    Such a body evolves  classically,   with a well-defined trajectory,  obeying Newton's equations
  \cite{KK2}.

\subsection{The microscopic degrees of freedom inside a macroscopic body are  quantum  mechanical   \label{microsub}}

The present discussion on the quantum ratio is  concerned  with  the question 
how  classical behavior for the CM of a macroscopic body  emerges from QM.   An important fact to be kept in mind 
 is the following.  Even if a macroscopic  (or a mesoscopic) body, {\it as a whole}, might behave classically, due to environment-induced or self-induced (or thermal)  decoherence and due to its large mass \cite{KK2},  the internal microscopic degrees of freedom,  the electrons, the atomic nuclei, and the photons, 
remain quantum mechanical (see Sec.~\ref{EP}, Sec.~\ref{nuclei}, Sec.~\ref{Decoherence}).  All sorts of  quantum-mechanical processes  (e.g., tunnelling)  continue to be active  inside the body, even if 
various decoherence effects may be significant.
 These quantum phenomena constitute the essence of the physics of  polymers and of general macromolecules, therefore of biology.   
 They hold the key to  the answers to many questions in biology, genetics and in neuroscience, unanswered today (see for instance 
\cite{QuantumBiology,ChiaraM}).
  The consideration of the present note has nothing to add, directly, to these questions. 
  See however a few more related  general comments below, in  Sec.~\ref{Decoherence}.

\section{The quantum ratio illustrated}    

In this section the quantum ratio will be illustrated in several examples.

\subsection{Elementary particles \label{EP}} 

  The elementary particles  known today (as of the year 2024)  are the  quarks,  leptons (electron, muon, $\tau$ lepton),  the three types of neutrinos,  and the gauge bosons (the gluons, $W$, $Z$ bosons and the photon), with masses  listed below.  
\begin{table}[h] 
  \centering 
 \begin{tabular}{|c|c|c|c|c|c|}
   \rule{0pt}{3ex}  $2.16 $  ($u$)   &   $4.67$ ($d$)     &   $93.4$ ($s$)    &  $1.27  \cdot 10^3$  ($c$)  &    $4.18 \cdot 10^3$  ($b$)  &    $172.7   \cdot 10^3$  ($t$)    \\
\end{tabular} 
  \caption{\footnotesize    The quark masses in  {MeV}/{$c^2$};     the errors not indicated.   $1$ MeV$/c^2   \simeq  
  1.782661 \cdot
  10^{-27}$ g    
   }\label{quarkmass}
\end{table}

\begin{table}[h] 
  \centering 
  \begin{tabular}{|c|c|c|c|}
   \rule{0pt}{3ex}   $0.51099895$  ($e$)  &   $105.658$  ($\mu$)    &   $1776.86$  ($\tau$)  &  $m_{\nu} \ne 0$\;;\quad   $m_{\nu} < 0.8$ eV$/c^2$         \\
\end{tabular} 
  \caption{\footnotesize The lepton masses. The $e$, $\mu$ and $\tau$  masses given in MeV$/c^2$.       
   }\label{leptonmass}
\end{table}

\begin{table}[h]
  \centering 
  \begin{tabular}{|c|c|c|c|}
  \rule{0pt}{3ex}  photon  &  gluons   & ${W^{\pm}}$ (GeV$/c^2$)  &   ${Z}$ (GeV$/c^2$)    \\
  $0  $  &  $ 0$ &   $80.377\pm 0.012  $  &  $ 91.1876 \pm 0.0021 $       \\
  \end{tabular}
  \caption{\footnotesize Gauge bosons  and their masses  }
  \label{gbosons}
\end{table}

The fact that the processes involving these  particles 
  are very accurately described by the local quantum field theory  
   $    SU(3)_{QCD}\times  \{  SU(2)_L\times U(1)\}_{GWS}  $   
    up to the energy range of  $O(10)\,$TeV,     means  that
    \be   L_0   \lesssim  O(10^{-18})\, {\rm cm}\;.    \label{elempart}    \ee      
 In future, these  elementary particles might well turn out to be made of  some constituents unknown today, bound by some new forces yet to be discovered.  For the present-day 
 physics their size can be taken to be  \footnote{Virtual emission and absorption of a particle of mass $m$  gives 
a physical ``size", $ h / mc$, known as the Compton length,  to any quantum-mechanical particle.
This should however be distinguished from the size $L_0$  defined as the extension of its internal wave function.  }
 \be  L_0=0\;\qquad  .^.. \qquad   Q=\infty\;      \label{QuantumEP}.\ee
 {\it  The elementary particles are quantum mechanical. } 
 
 This notion is generally taken for granted by physicists,  even if no justification is (was) known, as such.   Here, as we are enquiring whether certain ``particle", be it an atom, molecule, 
macromolecule, a piece of crystal,  etc.,   behaves quantum-mechanically or classically, and under which conditions, 
  it perhaps makes sense to ask  {\it   whether or why }  an elementary particle is quantum mechanical \footnote{Any quantum particle such as electron 
  behaves ``classically"    under certain conditions   (the Ehrenfest theorem), e.g.,  when it is well localized,  and free or under homogeneous  electromagnetic field, and within the diffusion time.  This is however not what we mean by a classical particle.   
  }. 
  Introduction of the criterion of quantum ratio  
 offers an  immediate  (affirmative) answer to the first question, and explains the second. 
  By definition the elementary particles have no internal structures, hence no internal excitations.
  There is no sense in talking about their body-temperature or thermal decoherence.  
  
 The observation  that the elementary particles are quantum mechanical  in the light of quantum ratio  (\ref{elempart}), (\ref{QuantumEP}), might  sound  new, but  it is not really  so.  
 Actually,  it reflects the common understanding  matured around 1970  in high-energy physics community   (e.g.,   't Hooft, Cargese lecture \cite{tHooft}) that the law of 
 Nature at the microscopic level is expressed by a {\it  renormalizable, relativistic local gauge theory (a quantum field theory)}   of the elementary particles.  Such a theory describes  pointlike
 particles  ($L_0=0$) \footnote{Quantum gravity or string theory effects, possibly relevant near the Planckian energies   $M_{Pl}\sim 10^{19}  \, GeV$,  does contain a length scale $\sim 10^{-32} \, cm$.      
 It is beyond the scope of the present work to consider if and how these affect the discussion of quantum or classical physics at larger distances  ($\ge 10^{-18}$ cm)  we are concerned with here. 
   }.     
 
 The scale  or dilatation  invariance of this type of theories is broken  by the necessity of introducing an ultraviolet cutoff, a mass scale  $\Lambda_{UV}$, to regularize, renormalize and define a finite theory   (quantum anomaly).  Remarkably,  the scale invariance is restored by the introduction of
 the renormalized coupling constants, defined conveniently at some reference mass scale $\mu$, and giving them an appropriate $\mu$ dependence  (the renormalization-group equations). See e.g.,  Coleman's 1971 Erice lecture \cite{Coleman}. 
  
  The fixed length or mass scales  of our world,  mentioned in Sec.~\ref{size}, concern the infrared fate of such dilatation invariance. 
  These fixed 
   scales can be, ultimately, traced to  (i)   the vacuum expectation value $ \brc  \phi^0 \ckt \simeq   246 \,   {\rm GeV} $
 of the Higgs scalar field  in the  $SU(2)\times U(1)$  Glashow-Weinberg-Salam electroweak theory     
 and   (ii)  the  mass scale  
$ \Lambda_{QCD} \simeq 150  \,  {\rm MeV}\;,$  
dynamically generated by the strong interactions (Quantum Chromodynamics).  
They break the scale invariance.   All fixed length scales from the microscopic  to macroscopic world we live in  follow from them and from some dimensionless coupling constants in the  
   $SU(3)_{QCD} \times \{SU(2)\times U(1)\}_{GWS}$ 
  theory
   \footnote{For instance,  the nuclear size is typically of the order of the pion Compton length,  $h / m_{\pi} c$, and $m_{\pi}^2 \sim m_{u,d} \Lambda_{QCD}$.  The proton and neutron masses  ($\sim 940$  MeV/$c^2$)   are mainly given by  the strong-ineraction effects, $\sim  \Lambda_{QCD}$.       The Bohr radius is  $\hbar^2/ m_e e^2$.}.  

 \subsection{Hadrons and  Atomic nuclei   \label{nuclei}}  
          
The atomic nuclei, together with various hadrons (the mesons and baryons), are the smallest composite particles known today.
Until around 1960, the mesons ($\pi$, $K$, ...) and baryons ($p$, $n$,...)  used to be part of the list of   ``elementary particles",  together with leptons. As the theory of the strong interactions
(the quark model,  and subsequently, the quantum chromodynamics, a nonAbelian 
$SU(3)$ gauge theory of quarks and gluons) was established around 1974 - 1980,  they  were replaced by the quarks and gluons, as more fundamental constituents of Nature.  

 The atomic nuclei  are bound states of the nucleons, i.e., protons ($p$) and neutrons ($n$).  They are bound by the strong interactions 
and their size is of the order of  
\be  L_0\sim  A^{1/3}     \,\, {\rm fm}    \;,  \qquad  1\, {\rm fm}   \equiv   10^{-13} \,  {\rm cm}    = 10^{-5} \,\AA \;,  \label{fermi}    \ee
where $A$ is the mass number.
The Coulombic  wave functions in the atoms and molecules 
have extension  ($R_q$)  of the order of   $O(\AA)$, thus 
\be   Q=     \frac{R_q}{L_0}   \gtrsim     O(10^5) \;.  \label{atomic}  
\ee 
The atomic nuclei  are quantum mechanical.

To say that the atomic nuclei are quantum mechanical because of the atomic extension  (\ref{atomic}), is however certainly  too reductive. 
The atomic nuclei indeed may appear without being bound in atoms.  For instance the $\alpha$ particle is the nucleus of the helium atom,
but may come out of a metastable nucleus through the $\alpha$ decay,  and propagate as a free particle.   It possesses the size of the order of (\ref{fermi}), 
much larger than the typical size of an elementary particles,   (\ref{elempart}),  but    still for the processes typically involving the distance scales much  larger than $1$ fm,  it behaves as a pointlike particle, i.e.,  quantum mechanically,  just as any elementary particle.   Similarly for the proton,  the nucleus of the hydrogen atom, with $L_0 \sim 0.84  $ fm. 

%
%
%

\subsection{Stern-Gerlach experiment  \label{SG}} 

The next smallest composite particles known in Nature  are the atoms. They are Coulombic composite states
made of the electrons moving around the positively charged atomic nuclei,  almost pointlike (at the atomic scales) and $O(10^3\sim10^5)$ times heavier than the electron. 
%

Let us consider the well-known Stern-Gerlach process of atoms with a magnetic moment  in an inhomogeneous magnetic field.  
To be concrete, we take as an example  the very original Stern-Gerlach experiment with the silver atom \cite{SG},  
$Ag$,  
with mass  and size,
\be   m_{Ag} \simeq  1.79  \cdot  10^{-22} \, {\rm g}\;, \qquad   L_0  \simeq   1.44  \,   \AA\;. 
\ee
Having the electronic configuration  and the global quantum number {\footnote{ $[Kr]$ indicates the zero angular momentum-spin ($L=S=0$) closed shell of the Kripton electronic configuration describing the first  $36$ electrons.},    
\be        [Kr] \,  4d^{10} \, 5 s^1\;,  \qquad    {\!}^2 S_{1/2} \;. 
\ee 
the magnetic moment  of the atom is dominated by the spin of  the outmost electron, 
\be    {\boldsymbol  \mu} =    \frac{ e \hbar }{2 m_e c} \, g \,  {\mathbf s}\;,     
\ee  
where $g$ is the gyromagnetic ratio $g \simeq 2$ of the electron.  

The question is whether the silver atom, which is certainly a quantum-mechanical bound state of $47$ electrons, $47$ protons and $51$ neutrons and 
with mass $\sim 100$ times that of the hydrogen atom,  behaves  as a whole  (i.e.,  its CM)  as a QM particle with spin $1/2$ or a classical particle of magnetic moment   $\mu$. 

The beam of $Ag$  is  sent into the region, long $3.5$ cm,  of  an inhomogeneous magnetic field  with  $\partial B_z /\partial z \ne 0$,  as it proceeds  in the e.g., ${\hat x}$ direction.   The beam width, which reflects the apertures of the two slits used to prepare the well  collimated beam,  is about  $\sim 0.02-0.03$  mm wide \footnote{The transverse wave packet size  of the atom can be taken to be of this order.  The silver atom, having the mass roughly $100$  times  that of the hydrogen atom,  has the diffusion time of the order of $10^{-2}$  sec  (see Table~\ref{diffusion}), so that  the  diffusion during the travel of  $3.5$ cm  is entirely negligible. 
  }. After passing the region of the magnetic field,  the image of the atoms on the glass screen shows two bands clearly separated about $0.2$ mm in the direction of $\hat z$.     In other words, at the end of the region with the magnetic field, the atom is described by a split wave packet of the form,
\be   \psi =    \psi_1({\bf r}) |\!\uparrow\ckt+  \psi_2({\bf r}) |\!\downarrow\ckt  \,,   \label{wavepackets} 
 \ee
 with the centers of the  two subpackets  $\psi_1$ and $\psi_2$  at  ${\bf r}= {\bf r}_1$ and    ${\bf r}= {\bf r}_2$, respectively, 
 where    $ |z_1- z_2| \simeq  0.2$ mm. The spatial support of the wave function $\psi$ can be taken as 
 about that size.    It follows that 
\be     Q =    \frac{R_q}{L_0} \gtrsim    \frac{0.2 \, mm}{1.4 \, \AA} \simeq  10^6  \gg 1\,: \label{QR1}
\ee
the $Ag$ atom, as a whole,  is behaving perfectly as a quantum-mechanical particle. 

Actually, the fact that the 
wave packets are divided in two by an inhomogeneous magnetic field does not necessarily mean that the system is in a pure state of the form
(\ref{wavepackets}).   The wave function of the form (\ref{wavepackets}) corresponds to a $100$\%   polarized beam,  where all incident atoms are in the  same  spin state 
\be     c_1  |\uparrow\ckt + c_2  |\downarrow\ckt  =  \left(\begin{array}{c}c_1 \\c_2\end{array}\right)\;, \qquad  |c_1|^2+ |c_2|^2=1\;. \label{pure} 
\ee
If the beam is partially polarized or unpolarized,  the spin state is described by a density matrix $\rho$.  The pure state  (\ref{pure}) corresponds to  the density matrix
\be    \rho^{(pure)} =  \left(\begin{array}{cc}|c_1|^2 & c_1 c_2^* \\c_2 c_1^* & |c_2|^2\end{array}\right)\;,  
\ee
whereas a general mixed state is  described by a generic, Hermitian $2\times 2$  matrix  $\rho$  with
\be   \Tr \, \rho =  1\;, \qquad   \rho_{ii}  \ge 0\;, \qquad i=1,2\;.
\ee
In an unpolarized beam, $\rho=  \tfrac{1}{2}  {\mathbf 1}$.

What the Stern-Gerlach  experiment measures is the relative frequency that the atom arriving at the screen happens to have spin $s_z=  \tfrac{1}{2}$ or 
 spin  $s_z=  -\tfrac{1}{2}$.  Let 
 \be   \Pi_{\uparrow} =  |\uparrow\ckt\brc \uparrow|  = \left(\begin{array}{cc}1 & 0 \\0 & 0\end{array}\right) \;, \qquad    \Pi_{\downarrow} =  |\downarrow\ckt\brc \downarrow|    = \left(\begin{array}{cc}0 & 0 \\0 & 1\end{array}\right)   \ee
 be the projection operators on    the spin up (down) states;    
 the relative intensities of the upper and lower blots on the screen are,   according to QM,  
\be    {\bar \Pi_{\uparrow} } =  \Tr   \Pi_{\uparrow}  \rho =  \rho_{11}\;;  \quad    {\bar \Pi_{\downarrow} } =  \Tr   \Pi_{\downarrow}  \rho =   \rho_{22}\;, \qquad 
 \rho_{11}\ +  \rho_{22}=1\;.    \label{density}     \ee
The prediction about the relative intensities of the two narrow atomic image bands  from the wave function  (\ref{wavepackets}) and from the density matrix, 
(\ref{density}) is in general indistinguishable, as is well known.  
In other words,  what the Stern-Gerlach experiment has shown is not whether the  atom is in a pure state of the form,  (\ref{wavepackets}), or in a generic  (spin-) mixed state, (\ref{density}),  
but that  the silver atom is a quantum-mechanical particle.  

Indeed,  the prediction for a classical particle is qualitatively different.  Each atom, if classical,  would  move    depending on the orientation of its magnetic moment,  tracing a well-defined trajectory, 
\be   m\, {\dot {\bf r}} = {\bf p}\;, \qquad       \frac{d {\bf  p}}{dt} =    {\bf F}  =   \nabla   ( {\boldsymbol {\mu}  \cdot {\bf B}})\;. \label{Newton} 
\ee
It would arrive at some  generic point on the screen.  If the initial orientation of the magnetic moment is random,  after many classical atoms have arrived they would leave a continuous band of atomic images,  instead of two narrow, well separated bands,  as has been experimentally observed  and as QM predicts.  

Vice versa, if 
  the orientation of the magnetic moment / spin is fixed (and the same)  for all incident atoms,  
  the classical atoms will produce only one narrow band, whereas QM atoms will leave   two separate  image bands.   
 These considerations clearly tell that the concepts of mixed state and classical particle should be distinguished.  We extend these discussions further   in  Sec.~\ref{Decoherence}, taking into account also the effects of environment-induced decoherence,  as well as the large spin limit, and their relations to the classical limit,   (\ref{Newton}).

\subsection{Atomic and molecular interferometry   \label{Interferometry}}

Many beautiful  experiments exhibiting  the quantum-mechanical feature (wave character)  of atoms and molecules  have been performed (or proposed), by using various types of  interferometers  \cite{Keith}-\cite{Bateman}.  One of the most powerful approaches uses the  Talbot-Lau  interferometry  \cite{Keith,C70Bis,C70,BrezArndtZeil,Chapman}.

The essential part of all these experiments makes use of the so-called Talbot effect \cite{Talbot}. 
In a typical setting, an atomic or molecular  beam,  passing through the first slit is sent to the diffraction grating  ($G_2$ in  Fig.~\ref{Talbotimage}), consisting of many 
slit apertures set with period $d$. 
After the passage of the diffraction grating,   the wave function of the atom or molecule, which originated from a point source,   has the form,   
\be       \psi(x_2)    \simeq    \sum_i    \psi_i(x_2)\;,      \label{aftergrating}
\ee
where   $  \psi_i(x_2)$  is the (transverse) wave packet of the atom (molecule)  which has passed through the $i$-th slit.   
 Just behind  the diffraction grating, the distribution of the atom (molecule)  has  a modulation such  as in Fig.~\ref{modulation}, each peak corresponding to the position of a slit opening.

 \begin{figure}
\begin{center}
\includegraphics[width=5in]{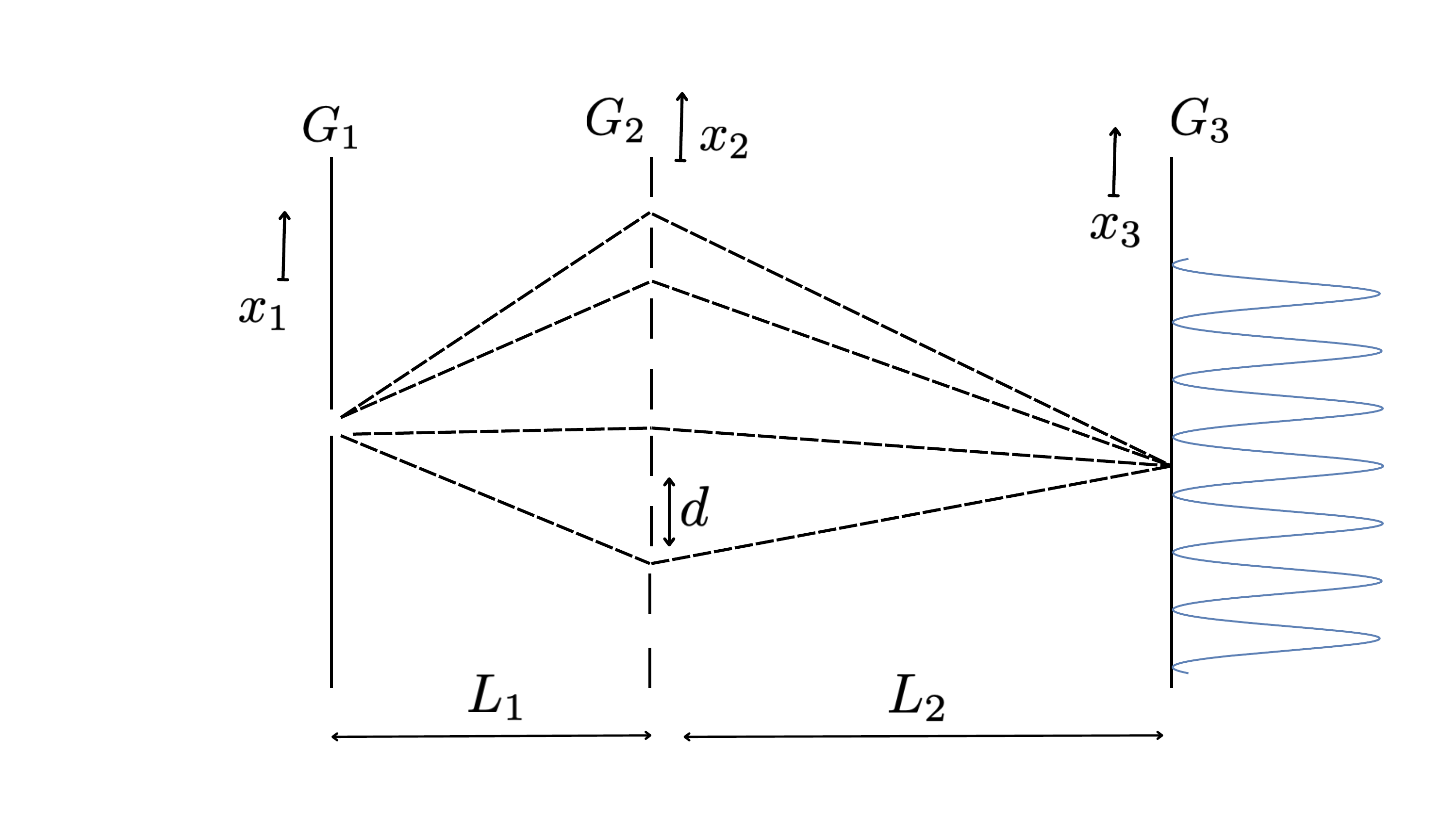}
\caption{\footnotesize The Talbot effect.  The intensity modulation of the molecules  immediately after the passage of the diffraction grating $G_2$, Fig.~\ref{modulation}, is reproduced,  due to the sum over paths,  at an imaging plane $G_3$ placed at definite distances  $L_2$, related to the Talbot length  (\ref{TalbotL}), from $G_2$.   }
\label{Talbotimage}  
\end{center}
\end{figure}

 \begin{figure}
\begin{center}
\includegraphics[width=5in]{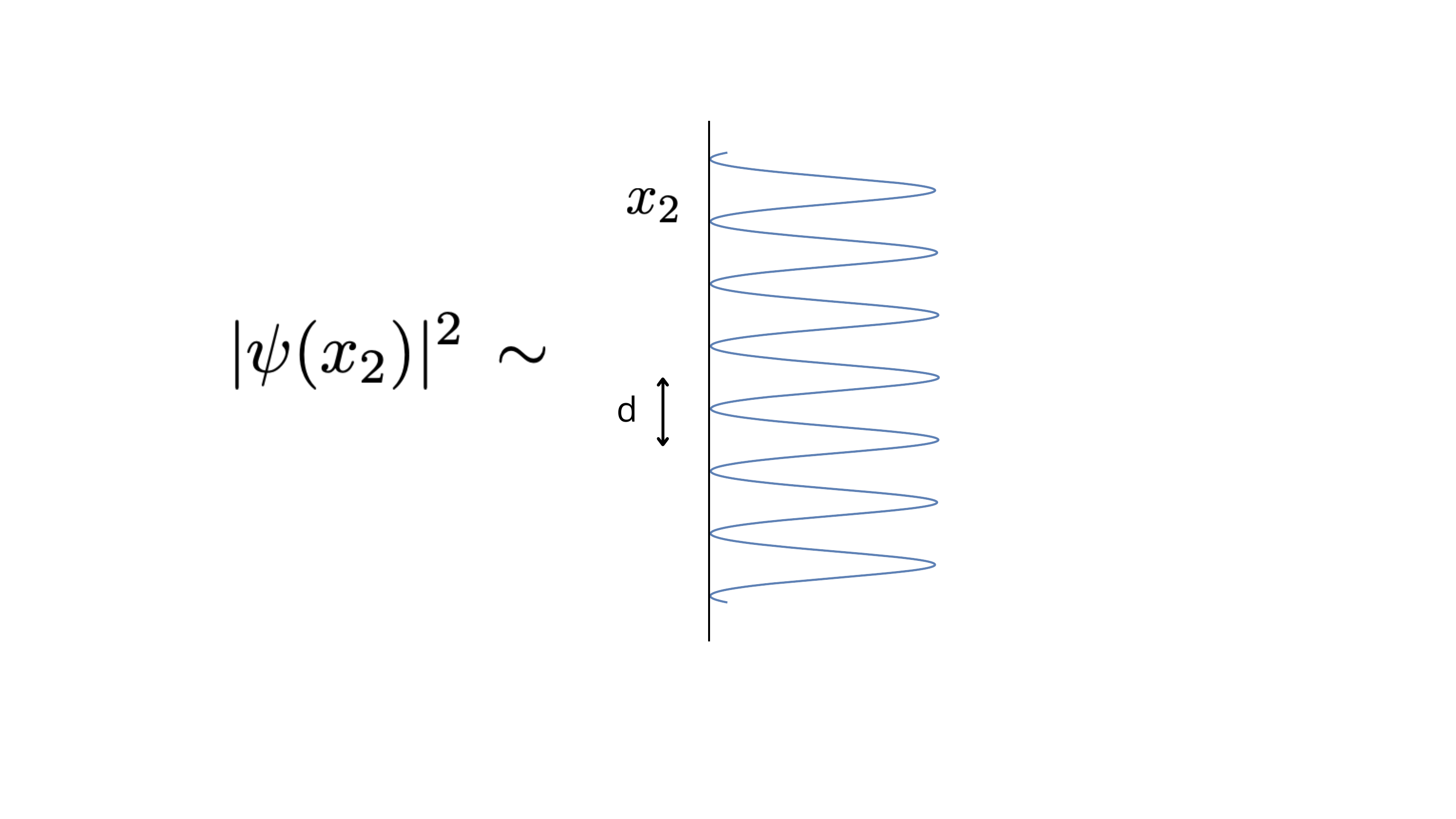}
\caption{\footnotesize   The intensity pattern of the atom (or molecule) behind the diffraction grating  $G_2$.  Each peak corresponds to a slit opening.  }
\label{modulation}
\end{center}
\end{figure}

 The coherent  components of the wave function,  (\ref{aftergrating}),  corresponding to different paths shown in Fig.~\ref{Talbotimage},   interfere constructively or destructively, depending on the vertical position $x_3$  on the imaging screen,  set at a distance $L_2$  from the diffraction grating. 
  In the so-called near-field diffraction - interference effects,
  the intensity modulation behind the diffraction grating  (Fig.~\ref{modulation})   is reproduced (self-imaging)  on the screen  $G_3$   \cite{Talbot,Keith,C70Bis,C70,BrezArndtZeil,Chapman}, 
 when  $L_2$   takes definite values, related to the Talbot length   \footnote{ $d$ is the period  of the slits in the diffraction grating,  and  
$  \lambda_{dB}  = \frac{h}{p}  $   
is the de Broglie wavelength ($p$ is the longitudinal momentum)   of the atom or molecule.}, 
\be     L_{T}  =    \frac{d^2}{\lambda_{dB}}\;, \label{TalbotL}
\ee  
as shown in  Fig.~\ref{Talbotimage}. 
The details of the calculation on the sum over paths  can be found in \cite{BrezArndtZeil}.  By introducing the geometrical magnification factors,   
$M_1 \equiv  (L_1+L_2)/L_2$;   $M_2 \equiv  (L_1+L_2)/L_1$, the sum over the  different paths of Fig.\ref{Talbotimage}     is shown to  give, for instance at $L_2=  2 M_2  L_T$,    the intensity pattern $|\psi(x)|^2$ behind the diffraction grating   reproduced at  $G_3$,  with an enlarged period, $M_2  d$. 
For   $L_2=   M_2  L_T$, instead, the same intensity pattern appears but shifted by a half period,  $x_3 \to x_3 + M_2 d /2$.

 In the Talbot-Lau interferometer, which is a variation of the above,  the imaging screen is replaced by a vertically  (i.e., in the $x_3$ direction) movable transmission-scanning  grating, with an appropriate  period $d^{\prime}$.  See Fig.~\ref{TalbotLau}.  This way, the  occurrence of the  interference fringes   - the Talbot  self-imaging -  is converted to the total transmission rate of the molecules (atoms) which varies periodically as a function of the vertical   ($x_3$)  position of the scanning grating  $G_3$ as a whole.  Another advantage of the   Talbot-Lau interferometer   is the possibility of introducing the incoherent source beam hitting the first grating. Even though the coherence sum over paths is relevant only for the atoms (or the molecules) which have originated from a definite source slit,  the use of an incoherent source can increase the total counts after  the third,  transmission grating, improving significantly the statistics of the experiments.

 \begin{figure}
\begin{center}
\includegraphics[width=5in]{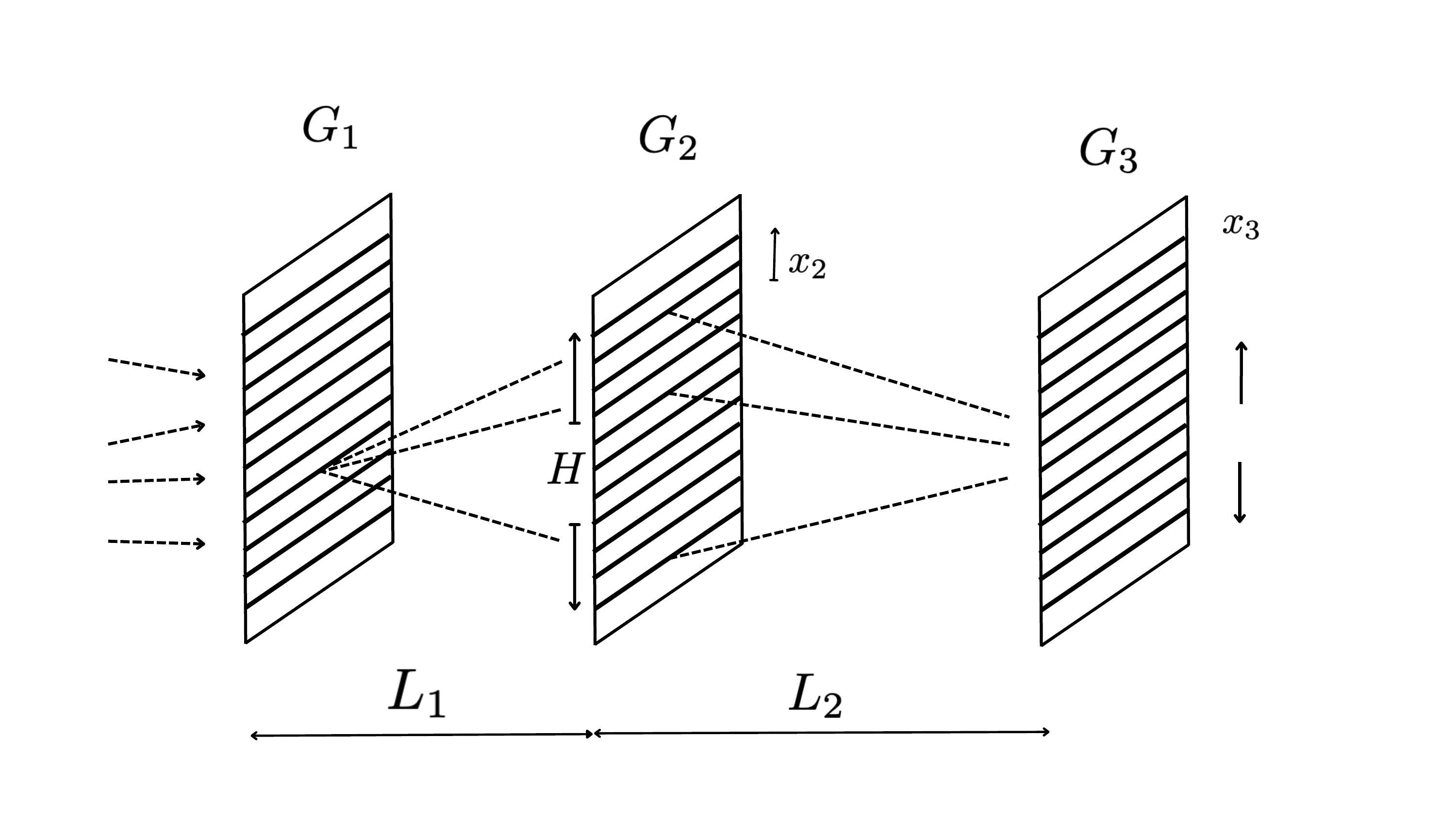}
\caption{\footnotesize   Talbot-Lau interferometer.  $G_2$ is the diffraction grating. Thick lines are the slit openings.  $G_3$ is a transmission-scanning grating movable vertically.   $G_1$ are  the  source slits.  }
\label{TalbotLau}
\end{center}
\end{figure}

For the purpose of discussing the quantum-fluctuation range and the quantum ratio,  these details of the set-up
  are  not really  fundamental.  
We need  simply to know the transverse, spatial extension of the wave function, (\ref{aftergrating}).  This  in turn can be taken as the height of the diffraction grating $G_2$, $H$,  of Fig.~\ref{TalbotLau}.    The detection of the Talbot effect or Talbot-Lau  fringe visibility,  is a proof that the transverse wave packets in (\ref{aftergrating})  are indeed in coherent superposition: i.e., that it is a pure state.  We thus take the quantum fluctuation range  $R_q$ 
in Table~\ref{QuantumRatio}   from the experimental total height  $H$ of the diffraction grating  $G_2$,  Fig.~\ref{TalbotLau}.   

A large quantum ratio implied by such a quantum range (and their size) is certainly an indication that these atoms and molecules are quantum mechanical,  
even though such an observation does not supersede the direct evidence of quantum coherence and interference effects presented in  \cite{Keith,C70Bis,C70,BrezArndtZeil,Chapman}.

\begin{table}  \centering 
  \begin{tabular}{|c|c|c|c|c|c|c|}
\hline
Particle &  mass    &    $L_0$  &    $R_q$   &    $Q $ &   Exp  &   Miscl  \\
\hline
  $Ag$   &  $108 $     &   $1.44 $   &    $0.2$  &    $\sim 10^6$  &      \cite{SG}   &   Stern-Gerlach   \\  
  $Na$   &   $23 $  &  $2.27$  &    $0. 5 / 0.75 $  &   $\sim 10^6$  &      \cite{Chapman}     &  \\  
      $C_{70}$   &  $840$    &      $9.4$       &    $16$    &    $\sim  10^7$  &     \cite{C70Bis,BrezArndtZeil,C70}  &       $T \ll 2000K$       \\     
            $C_{70}$   &  $840$    &        $9.4$      &    $\sim  0.001$        &    $ \sim  10^3  $        &    \cite{C70}   &     $T \ge 3000K$   \\                
\hline
\end{tabular}
  \caption{\footnotesize  The  size ($L_0$), the quantum fluctuation range ($R_q$)   and the 
  quantum ratio     $Q \equiv  R_q/L_0$   of atoms and molecules   in various  experiments.  The mass is in atomic unit  ($au$);  $L_0$ is given in Angstr\"om  ($\AA$);   $R_q$  is in $mm$.  In all cases,  the momentum of the atom (molecule), their masses, the size of the whole experimental apparatus, thus the time interval involved, 
are such that the  quantum diffusion of their (transverse) wave packets are negligible.
      }  \label{QuantumRatio} 
\end{table}

\subsubsection{On the ``matter wave"}

A familiar word used in the articles on the atomic and molecular interferometry  \cite{Brand}-\cite{Bateman}  is  ``matter wave".  
It might appear to summarize well the characteristic feature of quantum-mechanics:  ``wave-particle duality". 
Actually,   such an expression is more likely to obscure the essential quantum mechanical features of these processes, rather than clarifying them.  It appears to imply that the beams of atoms or molecules  somehow behave as a sort of wave:  that is not quite an accurate description of the processes studied  in   \cite{Brand}-\cite{Bateman}.   

 The wave-particle duality of de Broglie, the core concept of quantum mechanics, refers to the property of  {\it  each single quantum-mechanical particle,}  and not  to any unspecified  collective motion of particles in the beam  \footnote{The ``wave nature" of atoms or molecules 
 observed in  the  interferometry  \cite{Brand}-\cite{Bateman}  must   be distinguished from  the  many-body collective quantum phenomena, such as Bose-Einstein condensed ultra cold atoms described by a macroscopic wave function.}.
  This point was demonstrated experimentally by Tonomura et. al.\cite{Tonomura} in a double-slit electron  interferometry experiment   \`a la Young,  with exemplary clarity.   

 Exactly the same phenomena occur in any atomic or molecular interferometry.  For each single incident atom or molecule,  it amounts to the position measurement at the third,   
 imaging screen.
   For each incident  particle,   the result for the  exact final vertical position 
at   $G_3$  is not known:   it cannot be predicted,  in accordance with QM. 
 Only after the data with many incident particles are collected,  one  observes the interference effect  reflecting the coherence among the components of the extended wave function, (\ref{aftergrating}), 
 in accordance with the QM laws. 

    From the data given in   \cite{Brand}-\cite{Bateman}  it is not difficult to verify that the average distance 
between the successive atoms (or the molecules),   as compared to the size of their longitudinal wave-packet (which can be  deduced from the momentum 
uncertainty $\Delta p$),  is many orders of magnitudes larger.   For instance, in the  case of  the sodium-atom experiment  \cite{Chapman},   their ratio   is
about  $6  \,cm /  47 \AA \sim   10^{7}$. In the case of the $C_{70}$  \cite{C70}  this ratio seems to be even greater.     The atoms or molecules  do arrive one by one.

As the correlation among the  atoms or molecules  in the beam is  negligible   (as  it should be),  and the position of each final atom/molecule is apparently random,  the resulting interference fringes  such as manifested in  the Talbot (or the Talbot-Lau) interferometers, is all the more surprising and interesting.  What these experiments  show  goes much deeper  into the heart of QM,   
 than  the  words,  ``matter wave" or ``wave-particle duality",  might suggest.

\section{Decoherence versus classicality   \label{Decoherence}}
The atomic and molecular experiments discussed  in Sec.~\ref{SG}, Sec.~\ref{Interferometry}   are all performed in a high-quality  vacuum  \cite{SG}-\cite{Bateman}.  This is necessary, lest  the scattering of the atom or molecule under study  with the environment particles, e.g., air molecules, destroy their pure quantum-state nature and let them lose the ability of exhibiting typical quantum phenomena such as diffraction, coherent superposition, and interferences.  These processes are known as the environment-induced decoherence \cite{Joos1}-\cite{Zurek2}.   Under
environment-induced decoherence  the object under consideration becomes a mixture.  Diffraction,  coherent superposition, and interference phenomena 
typical of pure quantum states   get lost. 

  {\it    But it 
does not necessarily mean that the system becomes classical. }     Being in a mixed state  is necessary for the system to behave classically, but is in general
 not sufficient \cite{KK2}.   
  Unfortunately, there seems to be a widespread and inappropriate  identification in the literature of these two concepts: mixed (decohered) states and classical states. 
 
 Consider the free  electron.  Its decoherence rate /  time,  has been  studied 
under various types of environments  \cite{Joos1}-\cite{Zurek2}. 
 For instance,  in the $300K$  atmosphere at $1$ atm pressure,  a free electron  decoheres in $10^{-13}$ s  \cite{Tegmark}.  
  But when it interacts subsequently  with
  other systems,  it does so quantum mechanically, not as a classical particle.  Once it comes out of the region with ``environment",  it emerges as a free particle,  a pure quantum state. The same can be said of the photon  (as of any other elementary particle).
  
A related remark may be made about  the  cosmic rays. The gamma ray (photons), neutrinos, protons, etc., which are produced in hot and dense interiors of stars,  once out, travel in cosmic space (a good  approximation of vacuum)  as free,  quantum particles in pure states.  
  
 In the experiment of \cite{C70},   $C_{70}$  molecules are excited by a laser beam, before they enter the interferometry. When the average temperature of the molecules exceeds $3000$K,  the Talbot-Lau interference fringe signals are found to 
 disappear, showing that the molecules became mixed states, in agreement with the decoherence theory \footnote{ Here decoherence is caused by  the excitation of the molecules and the ensuing photon emission. It is more appropriate to talk about thermal (or self-) decoherence  \cite{C70},\cite{Hansen}, rather than environment-induced decoherence.}.  The quantum fluctuation range $R_q$   becomes of the order of the diffraction grating period, $d$.  This value is given in Table~\ref{QuantumRatio}.   But this does not mean  that the $C_{70}$ molecules have become classical:   what one can conclude from \cite{C70}   is that the thermal decoherence renders the molecule incoherent, mixed state.

 Below we consider two more test cases  where the difference among the pure state,  decohered mixed state, and the classical state,   is
 seen very clearly.   
These  considerations  can  have far-reaching consequences.  
For instance, they may indicate a way out of the ``no-go"  verdict 
 for the relevance of  quantum mechanics in the brain dynamics  \cite{TegmarkBis}. They may be fundamental  in all microscopic processes underlying  biological systems  (see Sec.~\ref{microsub})  \cite{QuantumBiology,ChiaraM}.

 \subsection{Stern-Gerlach set-up, decoherence and classical limit}

The original Stern-Gerlach (SG) process was discussed already (Sec.~\ref{SG}).  What the experimental result shows is that the silver atom behaves as a quantum-mechanical particle, either in a pure or (spin-) mixed state.

Here we discuss the SG  process again,  in more detail,  in three different regimes,  (i)  a pure QM process;  (ii)  the environmental decoherence  (an incoherent, mixed state);
and (iii)  for a classical particle. The main aim of this discussion is to highlight  the differences between these different physics situations as sharply as possible.    

\subsubsection{Pure QM state   \label{SGpure}}

For definiteness let us take the incident atom with spin $s= \tfrac{1}{2}$   directed in a definite, but generic,  direction,   ${\bf n}=  (\sin \theta  \cos \phi,  \sin \theta  \sin \phi,   \cos \theta)$,  i.e.,  
\be  \Psi 
=   \psi({\bf r}, t)  \, |{\bf n} \ckt \;,
      \label{WF}
\ee
where  
\be      |{\bf n} \ckt    =    c_1 |\!\uparrow\ckt +  c_2 |\!\downarrow\ckt \;,   \qquad  c_1=   e^{-i \phi/2}  \cos \tfrac{\theta}{2}\;, \quad c_2=   e^{i \phi/2}  \sin \tfrac{\theta}{2}\;,    \label{WFspin12}  
\ee
and $ \psi({\bf r}, t)$ describes the wave packet of the atom, moving towards   the $\hat x$ direction,  before entering the region with an  inhomogeneous magnetic field  ${\bf B}$.      
 The Hamiltonian is given by
 \be   H=   \frac{{\bf p}^2}{2m}  + V\;, \qquad        V=  -  {\boldsymbol  {\mu}  \cdot {\bf B}}\;,\label{Hamiltonian}
\ee
\be    {\boldsymbol  \mu} =   \mu_B    \, g \,  {\mathbf s}\;,     \qquad \partial  B_z / \partial z \ne 0\;,  \label{Spin}
\ee
where $ \mu_{B} =  \frac{ e \hbar }{2 m_e c}$ is the Bohr magneton \footnote{ We recall the 
well-known fact that the gyromagnetic ratio $g \simeq 2$ of the electron and the spin magnitude $1/2$ approximately cancel, so $\mu_B$ 
is the magnetic moment of the atom such as $A_g$. 
      }
.  The time evolution of the system is described by the Schr\"odinger equation
\be    i \hbar \,  \de_t  \Psi  =   H\, \Psi  
\ee
where the total energy is conserved.  

An example of the  inhomogeneous magnetic field ${\bf B}$ appropriate for the Stern-Gerlach experiment is
\be      {\bf B}  =   (  0,   B_y,  B_z ), \qquad   B_y = - b_0 \, y, \quad B_z = B_0 +  b_0  \, z\;     \label{magneticfield}
\ee   
which satisfy   $\nabla \cdot  {\bf B} =  \nabla \times {\bf B} =0$.  The constant field $B_0$ in the $z$ direction is assumed to be large, 
\be   |B_0|  \gg       |b_0 \, y|\;.  \label{large}   
\ee
 in the relevant region of $(y,z)$  of the experiment.  
The wave function of the spin $1/2$ particle entering the SG magnet  has the form,  
\be    
  \Psi=  {\tilde \psi}_1({\bf r}, t) |\!\uparrow\ckt +   \, {\tilde \psi}_2({\bf r}, t)   |\!\downarrow\ckt \;.
\ee
By redefining the wave functions for  the upper and down  spin components  as
\be      {\tilde \psi}_1({\bf r}, t)  =   e^{i  \mu_B  B_0 t / \hbar}   \psi_1({\bf r}, t) \;, \qquad   {\tilde \psi}_2({\bf r}, t)  =   e^{-  i  \mu_B  B_0 t / \hbar}   \psi_2({\bf r}, t) \;,
\ee
   $\psi_{1,2}$ are seen to satisfy the separate  Schr\"odinger equations \cite{Platt} 
\be   i \hbar  \frac{\de}{\de t}   \psi_1 =     \left(    \frac{{\bf p}^2}{2m}   -      \mu_{B}  b_0    z    \right)       \psi_1 \;,\qquad  
  i \hbar  \frac{\de}{\de t}   \psi_2  =     \left(    \frac{{\bf p}^2}{2m}   +       \mu_{B}  b_0    z    \right)       \psi_2 \;.
     \label{SEq}
\ee
Note that the  mixed terms (involving $\psi_2$ on the right hand side of the first equation, $\psi_1$ in the second equation)  contain the coefficients with rapidly oscillating phase factors, 
\be      \mp i  \mu_B b_0  y\,  e^{\mp   2  i  \mu_B  B_0 t / \hbar}   \;
\ee
and vanish on the average. The condition (\ref{large}) is crucial here.   As explained in \cite{Astrom},  this can be interpreted classically as  the spin precession  effect around the large constant magnetic field $B_0 \hat z$, thanks to which  the forces on the particle in the transverse directions average out to zero.  The only significant force is in the $\pm \hat z$ direction due to the inhomogeneity of $B_z(z)$, which deflects the atom  
in that direction
\footnote{
With a magnetic field $B_0$ of the order of $10^3$ Gauss, 
the precession frequency is of the order of $10^{11}$ in the case of the original SG experiment with silver atom \cite{Astrom}.
The precession period is orders of magnitude shorter than the time the atom spends in the region with the magnetic field.}.

From (\ref{SEq}) and their complex conjugates,  the Ehrenfest theorems for spin-up and spin-down wave-packets  follow  separately,  
      \bea  &&   \frac{d}{dt}   \brc {\bf r}\ckt_1   =  \brc {\bf p}/{m} \ckt_1  \;,   \quad   
      \frac{d}{dt}  \brc {\bf p}\ckt_1  =   -    \brc  \nabla (\mu_{B}   B_z) \ckt_1  \;;      \label{New1}   \\
   &&    \frac{d}{dt}   \brc {\bf r}\ckt_2   =  \brc {\bf p}/{m} \ckt_2  \;,   \quad   
      \frac{d}{dt}  \brc {\bf p}\ckt_2  =   +  \brc  \nabla (\mu_{B}   B_z) \ckt_2  \;,   \label{New2}
\eea
where    $\brc {\bf r}\ckt_1  \equiv   \brc   \psi_1 | {\bf r} | \psi_1 \ckt$, etc.    
Namely, for a sufficiently compact initial wave packet,  $\psi({\bf r}, t)$, the expectation values of    ${\bf r}$ and ${\bf p}$ in
the up and down components   $\psi({\bf r}, t)_{1,2}$  follow    respectively  the classical trajectories of a spin-up or spin-down particle..    At the end of the region with the magnetic field ${\mathbf B}$   \footnote{We assume that the transit time of the whole process and  the mass of the atom, are such that the free quantum diffusion of the wave packets is negligible. 
See also Appendix~\ref{Variational}. } it is described as a split wave packet  of the form,   (\ref{wavepackets}), (\ref{QR1}).   

Note however that,  even though the two subpackets  might be well separated by a macroscopic distance  $|{\bf r}_1 -{\bf r}_2|$, they are still  in a coherent superposition. Its pure-state nature can be verified by reconverging them by using  a second magnetic field of opposite gradients, and studying their interferences  (the ``quantum eraser'' set-up).      

A  variational  solution of  (\ref{SEq})   is given in Appendix~\ref{Variational}.    
The splitting of the wavepacket is illustrated in Fig.~\ref{WSG} (a).

\begin{figure}
\begin{center}
\includegraphics[width=5in]{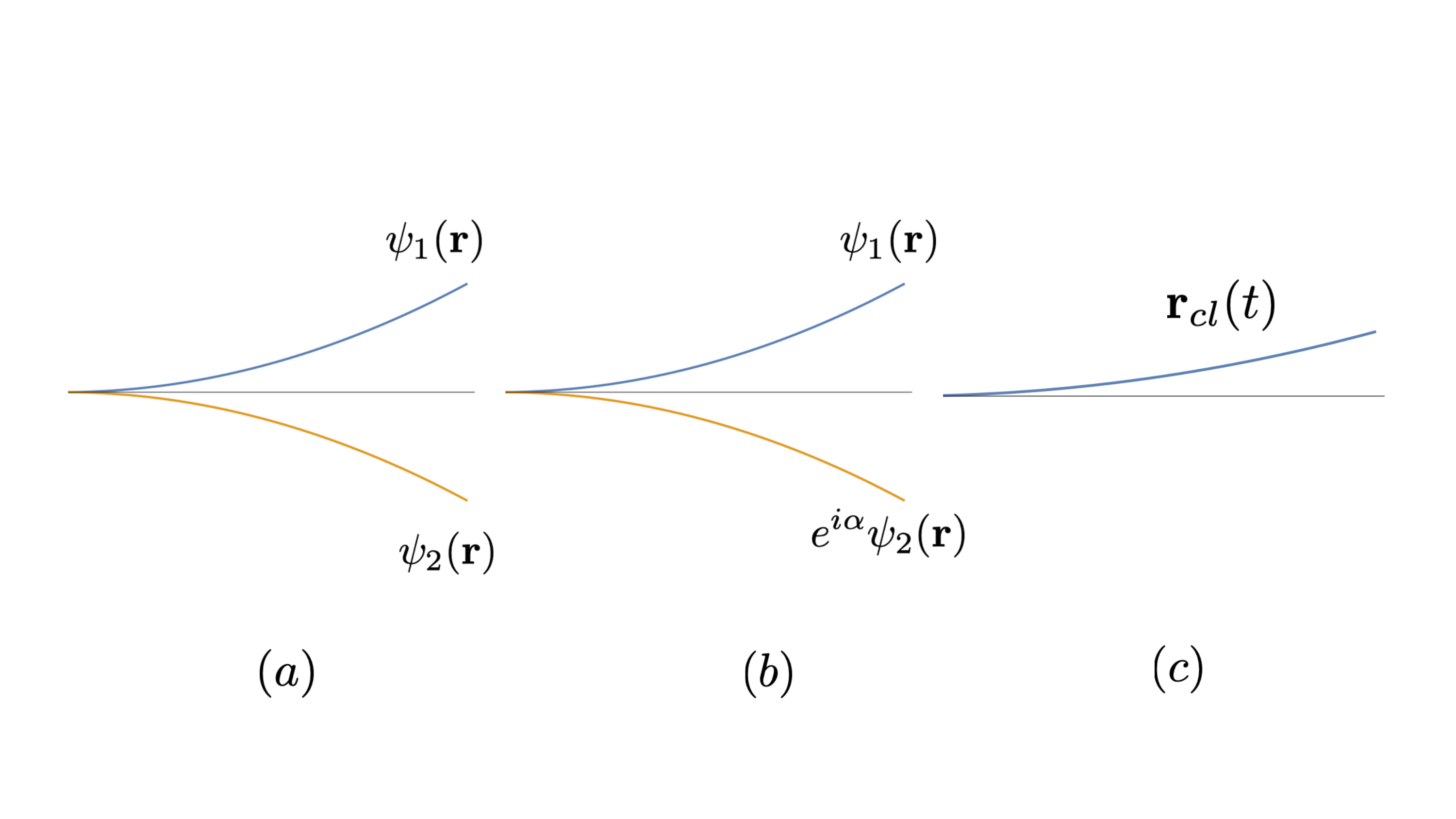}
\caption{\footnotesize  The spin-up and spin-down sub wavepackets of the silver atom evolve independently under the Schr\"odinger equation, both in the vacuum (fig.(a),  pure state)
and in a weak environment (\ref{cond1})-(\ref{cond3})   (fig.(b)),  where the decoherence is represented by an unknown, to-be-averaged-over,  relative phase $\alpha$ between $\psi_1({\mathbf r})$ and $\psi_2({\mathbf r})$.  Fig.(c)  represents a unique classical trajectory.     }
\label{WSG}
\end{center}
\end{figure}

  \subsubsection{Environment-induced decoherence     \label{SGmixed}} 
  
Let us now consider the SG  process  (\ref{WF})-(\ref{Spin})  again, but this time in a  poor vacuum,  e.g.,  in the presence of  non-negligible background. 
The decoherence of a well separated split wave packet as (\ref{wavepackets})  due to  the interactions with the  environment particles has been the subject of intense study  \cite{Joos1}-\cite{Zurek2}.        
The upshot of the  results of these investigations is that   the environment-induced decoherence causes the pure state   (\ref{wavepackets})   to become a mixed state   at $t\gg 1/\Lambda$,  described by a  diagonal  density matrix  
    \be \psi({\bf r})  \psi({\bf r}^{\prime})^*  \to    \psi_1({\bf r})     \psi_1({\bf r}^{\prime})^*  |\!\uparrow\ckt \brc \uparrow|    +   \psi_2({\bf r})     \psi_2({\bf r}^{\prime})^* 
    |\!\downarrow\ckt \brc \downarrow| \;,\qquad   |{\bf r}_1-{\bf  r}_2| \gg  \lambda\;,     \label{mixed} 
  \ee
 where $\Lambda$ is the decoherence rate  \cite{Joos1}-\cite{Zurek2} and $\lambda$ is the de Broglie wavelength of the environment particles.  The density matrix  (\ref{mixed}) means that each atom is now   {\it    either near ${\bf r}_1$ or  near   ${\bf r}_2$.}
   The prediction for the SG experiment is however  the same as the case of spin-mixed state
  (partially polarized source atoms), (\ref{density}):   it cannot be distinguished from the prediction $|c_1|^2 : |c_2|^2$ for the relative intensities of the two image bands, in the case of the pure state. 
  
 Actually, the study of the effects of the environment particles on any given quantum process  is  a complex, and highly nontrivial problem,   as it involves many factors, such as the density and flux of these particles,  the pressure, the average temperature,  kinds of the particles present and the type of interactions,  etc.   \cite{Joos1}-\cite{Zurek2}.    A simple statement such as (\ref{mixed}) might sound as   an oversimplification.

Without going into the detailed features of the environment,  we may nevertheless attempt to clarify the basic conditions under which the result such as  (\ref{mixed}) can be considered reliable.  
Following  \cite{Tegmark},  we  introduce  the {\it  decoherence time}  $\tau_{dec} \sim 1/ \Lambda_{dec}$,   as a typical timescale over which the decoherence takes place.  Also the {\it  dissipation time}  $\tau_{diss}$
may be considered, as a  timescale in which the loss of the energy, momentum  of the atom under study due to the interactions with the environmental particles, become significant  \footnote{
Unlike \cite{Tegmark}, however, we shall not consider $\tau_{dyn}$, the typical timescale of the internal motion of the object under study. 
Roughly speaking the size $L_0$ (the space support of the internal wave function) we introduced in defining  the quantum ratio,  (\ref{QR}),   corresponds to it ($\tau_{dyn}   \propto  L_0$).
Quantum-classical criteria   suggested by \cite{Tegmark} might appear to have some similarity with (\ref{quantum}), (\ref{classical}).  However,  the former seems to leave unanswered questions such as ``what happens to a quantum particle ($\tau_{dyn} < \tau_{dec}$), at  $t> \tau_{dec}$?"     This is precisely the  sort of question which we are trying to address here. 
}.  
We need to consider also a typical {\it  quantum diffusion time},   $\tau_{diff}$, and finally, the {\it  transition time},    $\tau_{trans}$,  the interval o f time the atom  spends between the source slit to the 
image screen (or anyway the final, reference position in the direction of motion).   

First of all, we assume that the velocity of the incident atom, its mass, and the size of the whole apparatus are such that  the free quantum diffusion  (the spreading of the wave packets)
is negligible,  during the process under study.  Furthermore,  the environment is assumed to be sufficiently weak so that  the effects of energy loss, the momentum transfer, etc. can also be neglected 
to a good approximation.   As  shown in    \cite{Joos1}-\cite{Zurek2}, the loss of the phase coherence is a much more rapid process than the dynamical effects affecting the motion of the particle considered.  

In other words, we consider the time scales 
\be    \tau_{dec} \ll  \tau_{trans}  \ll    \tau_{diff}, \tau_{diss}\;. \label{cond1}
\ee   
The first inequality tells that the motion of the wave packets is much slower than  the typical decoherence time.  Consider the atom at some point, where  it is described
by a split wave packet of the form (\ref{wavepackets}),  with their centers separated   by 
\be   |{\bf r}_1 - {\bf r}_2|  \gg  a\;,      \label{cond2}   \ee
   where $a$ is the size of the original wavepacket.  We may then treat such an atom as if it were  
 at rest,  and take into account the rapid decoherence processes studied in \cite{Joos1}-\cite{Zurek2} first  (a sort of Born-Oppenheimer approximation).
Furthermore,  let us  also take the environment particles  with a typical de Broglie wavelength  $\lambda$  such that 
\be     a \ll     \lambda    \ll       |{\bf r}_1 - {\bf r}_2| \;:   \label{cond3}
\ee
 the environment particle can resolve between the split wave packets,  but not the interior of  each of the subpackets,    $\psi_1({\bf r})$ or     $\psi_2({\bf r}).$ 

In conclusion, under the conditions  (\ref{cond1})-(\ref{cond3}),  each of the split wave packets proceeds just  as in the   pure case (no environment) reviewed in Sec.~\ref{pure}, 
whose average position and momentum  (i.e., the expectation values)  obey Newton's equations,   (\ref{New1}), (\ref{New2}).   Each of   the subpackets describes a quantum particle, in  a (position) mixed state,  that is, either near  ${\bf r}_1$
or  ${\bf r}_2$.   After leaving the region of the SG magnets, it is just a (pure-state) wave packet  $\psi_1({\bf r})$ or   $\psi_2({\bf r})$.  The  two possibilities no longer interfere, in contrast to the pure split wave packet studied in Sec.~\ref{pure}.   
See  Fig.~\ref{WSG} (b).

Needless to say,  if any of the conditions (\ref{cond1})-(\ref{cond3}) are violated,  the motion of the atom would be very different.  For instance,  $\tau_{diss} \ll  \tau_{trans}$   would mean a totally random motion for the atom.  Even in such a case,  though,  the effects of the environment-induced decoherence/disturbance are quite distinct from the motion of a classical particle, with a unique, smooth trajectory, 
discussed below.

  \subsubsection{Classical (or quantum?)  particle} 
  
A classical particle, with  the magnetic moment directed   towards    
\be  {\bf n}=  (\sin \theta  \cos \phi,  \sin \theta  \sin \phi,   \cos \theta)\;, \ee
   is described by  Newton's  equation, 
(\ref{Newton}).
The way  the unique trajectory for a classical particle emerges from quantum mechanics has been discussed in \cite{KK2},  where the magnetic moment is an expectation value 
\be          \sum_i   \langle \Psi |    ( {\hat   {\boldsymbol \mu}}_i +     \frac{e_i    {\hat  {\boldsymbol \ell}}_i  }{2  m_i  c } ) |\Psi \rangle  =   {\boldsymbol  \mu}\;,    \label{classicalSG}
\ee
taken in the internal bound-state wave function $\Psi$   and $\mu_i$ and    $\frac{e_i    {\bf \ell}_i  }{2  m_i  c }$ denote the intrinsic magnetic moment and one due to the orbital motion of the $i$-th constituent atom (molecule);    $i=1,2,\ldots, N$.  Clearly,   in general,  the considerations made in  Sec.~\ref{SGpure} and  Sec.~\ref{SGmixed} for a spin $1/2$ atom,  with a doubly split wave packet,    cannot be  generalized   simply to
(or compared with)  a classical body  (\ref{classicalSG})  with  $N \sim O(10^{23})$.

Nevertheless,  logically,  one cannot exclude particular systems  (e,g,   a magnetized metal piece)   with all spins directed in the same direction, for instance.  One might thus wonder how  a  quantum-mechanical particle of spin $S$   behaves  under an inhomogeneous magnetic field,   in the large spin limit,  $S=  \tfrac{N}{2}$, $N\to \infty$.  

The  question is   {\it     whether  
the conditions,   discussed in \cite{KK2},    for the emergence of classical mechanics  (with a unique trajectory)  for a macroscopic body (see also Sec.~\ref{Last} below),   are sufficient},  or  whether  some extra condition  or a mechanism  is needed to suppress 
possible, wide spreading  of the wave function into many subpackets  (see Fig.~\ref{Spreadsi})
under an inhomogeneous  magnetic field.

The answer is simple but somewhat unexpected.  Consider the state of  a  spin  $S$      directed towards  a direction,  $ {\mathbf  n}$, 
  \be    \left({\mathbf  S}\cdot {\mathbf  n} \right) \,  | {\mathbf  n}\ckt   = S\,  | {\mathbf  n}\ckt \; ,      \qquad  {\bf n}=  (\sin \theta  \cos \phi,  \sin \theta  \sin \phi,   \cos \theta) \;.  
\label{directed}   \ee 
Assuming that the magnetic field (and its gradients) is in the   ${z}$  direction,  we need to express  $ | {\mathbf  n}\ckt   $ as a superposition of the eigenstates of $S_z$,
\be     |{\mathbf  n}\ckt  = \sum_{k=0}^N        c_k \,   |S_z= M\ckt  \;, \qquad  M=   -  \frac{N}{2}  +  k \;, \quad (k=0,1,\ldots, N)\;.   \label{see} 
  \ee
 The expansion coefficients $c_k$ are known  \footnote{To get (\ref{excercise}),  consider  (\ref{directed})  as  
 a direct product  state  of   $N$ spin $\tfrac{1}{2}$ particles,      all oriented in the same direction, (\ref{WFspin12}).  Collecting terms with a fixed $k$ (the number of spin-up particles) gives 
 (\ref{excercise}).
    }:
 \be   c_k =      {\binom{N}{k}}^{1/2}   \,     e^{-i M \phi/2}\left(\cos\tfrac{\theta}{2}\right)^k   \left(\sin\tfrac{\theta}{2} \right)^{N-k} \;, \qquad   \sum_{k=0}^N  |c_k|^2 =1\;, \label{excercise}
\ee   
where  $ \binom{N}{k} $ are binomial coefficients, $N!/k!(N-k)!$. 
By using Stirling's formula,   one finds at large $N$ and $k$  with  $x= k/N$  fixed,     the distribution in  various $S_z=M$, 
\be |c_k|^2   \simeq    e^{N f(x)}\;, \qquad   x= k/N\;,    \ee
with
\be   f(x)   =  - x \log x - (1-x) \log (1-x)    + 2 x  \log  \cos \tfrac{\theta}{2}  + 2 (1-x) \log   \sin \tfrac{\theta}{2}  \;. 
\ee
The saddle-point approximation, valid at $N\to \infty$,  gives 
\be   f(x)    \simeq     - \frac{(x-x_0)^2 }{x_0(1-x_0)}   \;,  \qquad      x_0  =     \cos^2  \tfrac{\theta}{2}\;.      \label{spike1} 
\ee
and thus 
\be      |c_k|^2      \longrightarrow          \delta(x-x_0)    \label{spike2} 
\ee
in the $N\to \infty$  ($x= k/N$ fixed)  limit.
The narrow peak position $x=x_0$  corresponds  (see   (\ref{see}))   to  
\be  S_z =M =  N(x-\tfrac{1}{2}) =   S  \, (2  \cos^2  \tfrac{\theta}{2}-1)  =   S\, \cos \theta\;.  
\ee
Thus  a large  spin ($S\gg \hbar$) quantum particle with spin directed towards ${\mathbf n}$, in a  Stern-Gerlach setting with an inhomogeneous magnetic field, 
${\mathbf B} =  (0, 0, B(z))$,     moves  along a single trajectory 
of a classical particle  with  $S_z=   S\, \cos \theta$,  instead of spreading over a wide range of  split subpacket trajectories covering   
$-S \le  S_z   \le  S$!    See Fig.~\ref{Spreadsi} and Fig.~\ref{Spreadno}.

\begin{figure}
\begin{center}
\includegraphics[width=3.5 in]{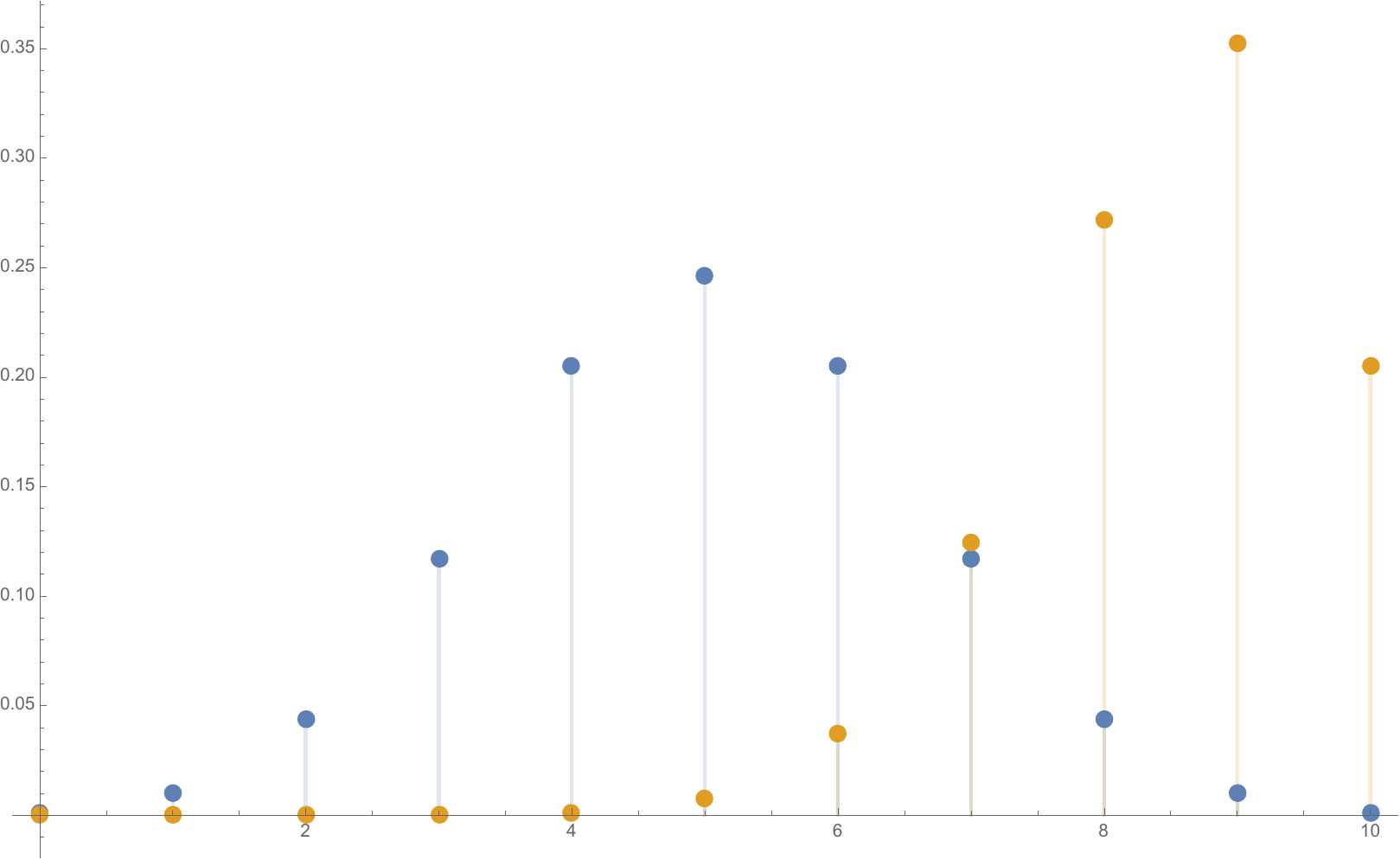}
\caption{\footnotesize {The distribution   $|c_k|^2$ in $k$, i.e., in possible values of $S_z$, $-S\le S_z\le S$,    for a spin $S=5$ particle  in the state  (\ref{directed}),   with $\theta = \pi/2$ (center) and 
$\theta = \pi/4$  (right).  }}
\label{Spreadsi}
\end{center}
\end{figure}

\begin{figure}
\begin{center}
\includegraphics[width=3.5 in]{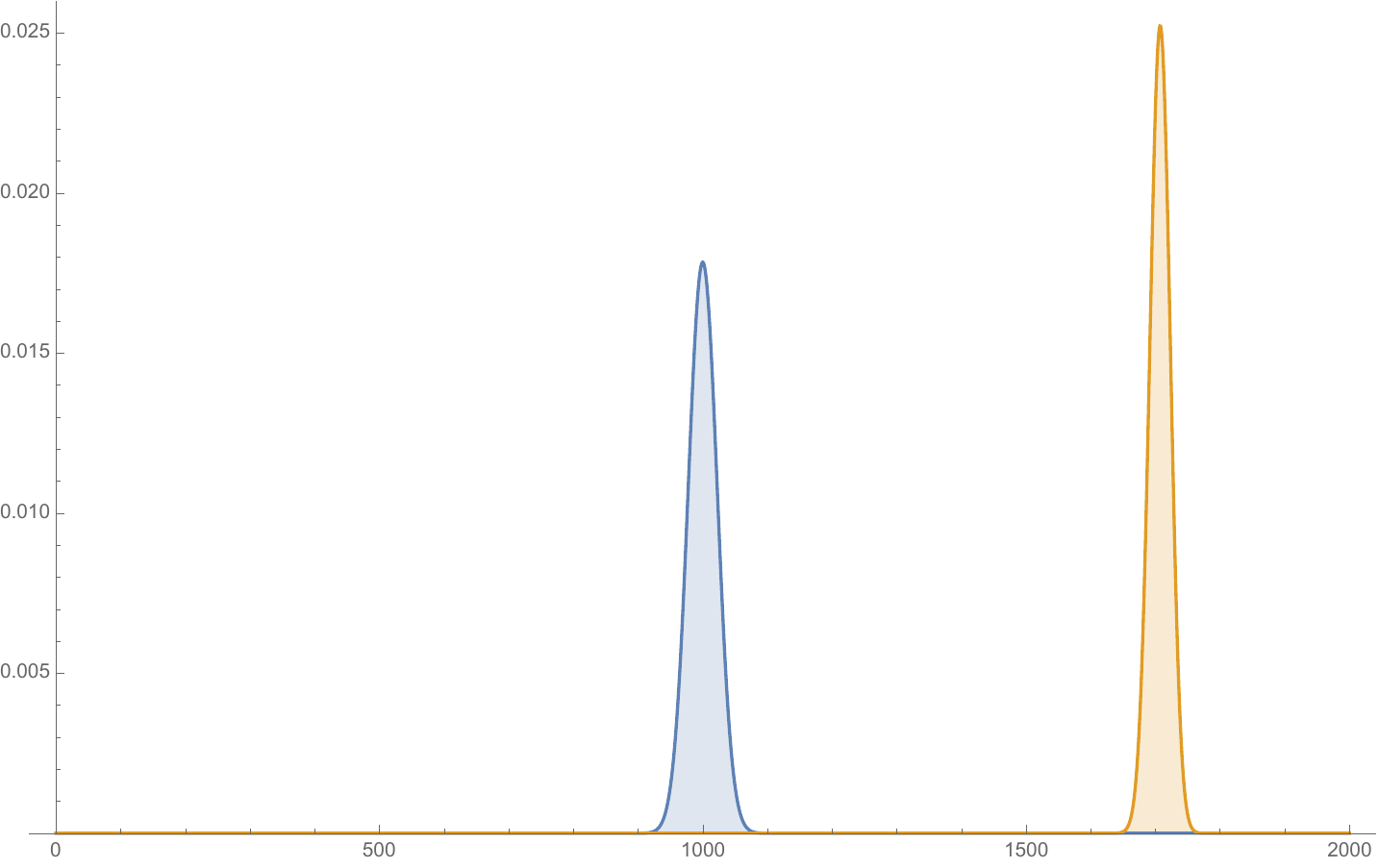}
\caption{\footnotesize {The distribution in possible values of $S_z$, for a spin $S=10^3 $ particle  in the state  (\ref{directed}),   with $\theta = \pi/2$ (center) and 
$\theta = \pi/4$  (right).  The particle starts looking like classical.  }  }
\label{Spreadno}
\end{center}
\end{figure}

This somewhat surprising result means that  QM  takes care of itself,  in showing that a large spin particle   ($S/\hbar \to \infty$)  follows a classical trajectory  \footnote{If the value  $S \gg \hbar $  is understood as due to the large number of spin $1/2$ particles composing it (see the previous footnote),  the spike  (\ref{spike1}),   (\ref{spike2}) can be  understood as due to the accumulation of an enormous number of microstates giving $S_z=M$.   
   }
consistently with the  known general behavior  of the wave function in the semi-classical limit  ($\hbar \to 0$)  \footnote{Of course this does not mean that the classical limit necessarily
requires or implies $S \to \infty$.}.

 \subsection{Tunnelling molecule}  

As another example,   let us consider a toy version of an atom (or a  molecule) of mass $m$  moving  in the $z$ direction with momentum $p_0$,   but with  a split wave packet in the transverse ($x,y$) plane,
 \be   \Psi   =    e^{i p_0  z / \hbar}   \psi(x, y)\;, \qquad     \psi(x,y) =   c_1  \psi_1(x,y) +    c_2  \psi_2(x,y)\;, \label{Trwv} 
 \ee
 where $\psi_1$ and $\psi_2$ are narrow (free) wave packets centered at   ${\bf r}_1=(x_1,y_1)$ and   ${\bf r}_2 =  (x_2,y_2)$, respectively.   
 This is somewhat analogous to  the wave function of  the silver atom  (\ref{wavepackets})  or  of  the $C_{70}$ molecule  (\ref{aftergrating}).   
 Actually,  we take also for  the longitudinal wave function a 
 wave packet,   $\chi_{p_0}(p, z)$, by considering  a linear superposition of the plane waves  $e^{i p   z / \hbar}$ with momentum $p$  narrowly distributed around  $p = p_0$.   For instance a Gaussian distribution in  $p$,   $\sim e^{- (p-p_0)^2 / b^2}$,   yields a Gaussian longitudinal wave packet in $z$ of width $ \sim 2 \hbar/b$. 
 At  times much smaller than the characteristic diffusion time    $t \ll   \frac{2m \hbar}{b^2}$    \footnote{
 The exact answer has  the Gaussian width in the exponent replaced as 
 $\tfrac{b^2}{4\hbar^2} \to   \tfrac{b^2}{4\hbar^2  ( 1 +   i b^2 t/ 2 m \hbar)}$, and the overall wave function multiplied by  $ (1 +  i \, b^2  t  / 2m \hbar)^{-1/2}$.  
 These are  the standard  diffusion effects of a free Gaussian wave packet of width  $a =  2 \hbar/b$.  
 Also,  if the longitudinal wave packet and the transverse subwave packets are taken to be of a similar size,  then the free diffusion of the transverse wave packets 
 (hence $t$-dependence of $\psi(x,y)$) can also be neglected.
 },  the particle is approximately described by the wave function,   
 \be     \Psi_{asymp}    \sim   e^{i p_0 z/\hbar}  e^{-  i p_0^2  t   / 2 m \hbar}  e^{- \tfrac{ b^2}{4\hbar^2} ( z -   \tfrac{p_0 t }{m})^2  }  \;    \psi(x, y)\;. 
 \ee

 Assume that such a particle is incident from $z= -\infty$   ($t=-\infty$), moves towards right (increasing $z$),   and hits a potential barrier    (Fig.~\ref{ET1}),
 \be       V   =     \begin{cases}
   0 \;,   &     |z|  >    a , \\
    V(z)   \;,   &    - a   <  z  <   a 
\end{cases}
 \ee
 whose height is above the energy of the particle, approximately given by the longitudinal kinetic energy, $E \simeq   \tfrac {p_0^2}{ 2 m}$.    
 As the longitudinal and transverse motions are factorized,   the relative frequencies (probabilities)  \footnote{It was proposed  in \cite{KK1,KKTalk}  to use ``(normalized)  relative frequency"  instead of the word  ``probability".  The traditional probabilistic Born rule places the human intervention as the central element of its formulation, and distorts the
way quantum-mechanical laws   (the laws of Nature!) look. To the authors' view, this was the origin of innumerable puzzles, apparent contradictions and conundrums  
entertained in the past.  See  \cite{KK1,KKTalk}   for a new perspective and a more natural understanding of the QM laws. 
}    of finding the particle on both sides of the barrier
 (barrier penetration and reflection) at large $t$  can be calculated  by the standard one-dimensional QM. The answer is well known:  the tunnelling frequency   is given, in the semi-classical approximation, by 
 \be   P_{tunnel} =  |c|^2\,,  \qquad  c  \sim   e^{ - \int_{-a_0}^{a_0}     dz  \, \sqrt{ 2m (V(z)-E)} / \hbar  }  \;,   \label{penetration}   
 \ee
  ($V(z)-E>0,\,\,   -a_0 < z < a_0$).  
   The particle 
  on the right  of the barrier  is described by the wave function
  \be    \Psi_{penetrated} \simeq    c\, \Psi_{asymp} =  c\,    e^{i p_0 z/\hbar}  e^{-  i p_0^2  t   / 2 m \hbar}  e^{- \tfrac{ b^2}{4\hbar^2} ( z -   \tfrac{p_0 t }{m})^2  }  \;    \psi(x, y)\;,  \label{right} 
  \ee
   where  $c$ is the transmission coefficient of 
   (\ref{penetration}).
  The transverse, coherent superposition of the two components  (sub  wavepackets), (\ref{Trwv}),  remains intact.   See Fig.~\ref{ET1}.
  
\begin{figure}
\begin{center}
\includegraphics[width=5in]{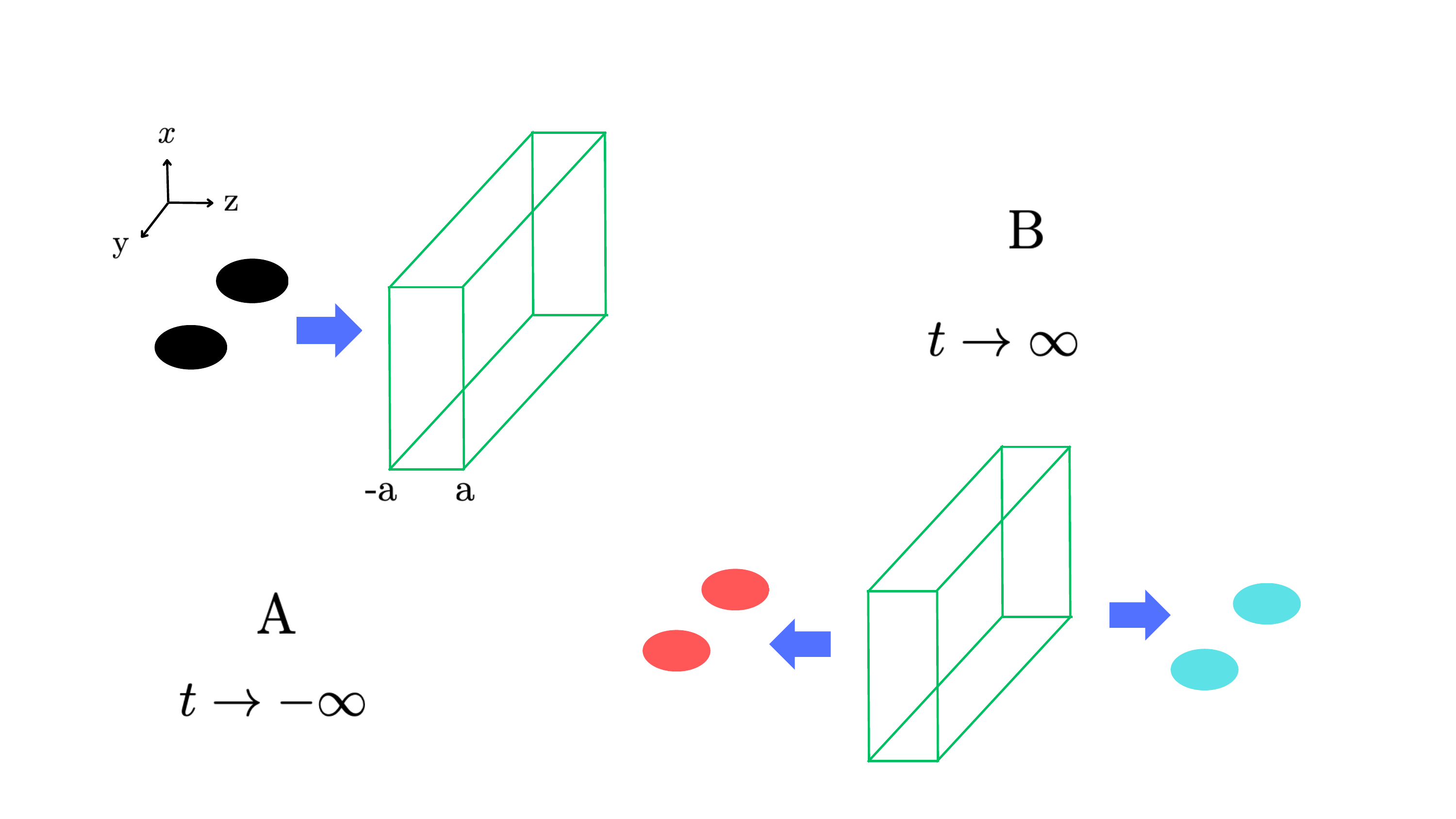}
\caption{\footnotesize  On the left figure (A),  an atom (molecule) arrives from $z=-\infty$ and moves towards the potential barrier   $V(z)$  at $ -a < z < a$  (independent of $x$ and $y$).   It is described by a wave packet (split in the transverse direction as in (\ref{Trwv})).   The wave function  of the particle   at $t \to \infty$,  shown in the right part (B),  contains both the reflected and transmitted waves. The coherent superposition of the two sub wavepackets in the $(xy)$ plane  remains intact.  
 }
\label{ET1}
\end{center}
\end{figure}

Now reconsider the whole process, with the region left of the barrier  ($z < - a$) immersed in air  \footnote{Or, as  in the $C_{70}$ experiments \cite{C70},  the incident  molecules may get bombarded by laser beams,  get excited, and emit photons,  before they reach the potential barrier.  }.  The precise decoherence rate depends on several parameters,  but  the incident particles get decohered  in a very short time in general,  as in (\ref{mixed})  \cite{Joos1}-\!\cite{Zurek2}.   The particle at the left of the barrier \footnote{We assume that the environment particles  (air molecules)  have energy much less than the  barrier height, so that they are
confined in the region left of the barrier.}   is now a mixture:  each atom (molecule)  is either near  ${\bf r}_1=(x_1,y_1)$ or ${\bf r}_2 =  (x_2,y_2)$ in the transverse plane. 
But when it hits the potential barrier it will tunnel through it, with the relative frequencies (\ref{penetration}), and will emerge on the other side of the barrier  as a free  particle.   It has the wave function, (\ref{right}),  with  $\psi(x,y)$ replaced by  $\psi_1(x,y)$,  with relative frequency  $|c_1|^2 / (|c_1|^2+ |c_2|^2)$, and by  $\psi_2(x,y)$, with frequency    $|c_2|^2 / (|c_1|^2+ |c_2|^2$).  It is a statistical mixture, but each is a pure quantum mechanical particle. See Fig.~\ref{ET2}.

Our discussion above assumes that the air molecules (the environment particles) are just energetic enough (their de Broglie wave length small enough) to resolve the transverse split wave packets   (see (\ref{mixed})), but are at the same time much less energetic than the longitudinal kinetic energy   $\tfrac {p_0^2}{ 2 m}$  and that  their flux is sufficiently small. In writing (\ref{right}) we assumed that the effects of the environment particles on the longitudinal wave packet  are small, even though the tunnel frequency may be somewhat   modified, as it is very sensitive to its energy.

Obviously, in a much warmer and denser environment the effects of the scatterings on our molecule would be more severe, and the tunnelling rate would become considerably smaller.  Even then, our atom (or molecule) remains quantum mechanical  \footnote{The situation is reminiscent of the $\alpha$ particle track  in a Wilson chamber.   $\alpha$ is scattered by atoms,  ionizing them  on the way,  but traces roughly a straight trajectory. 
When it arrives at the end of the chamber, it is just the same $\alpha$ particle.  It has not become classical.}.

\begin{figure}
\begin{center}
\includegraphics[width=5in]{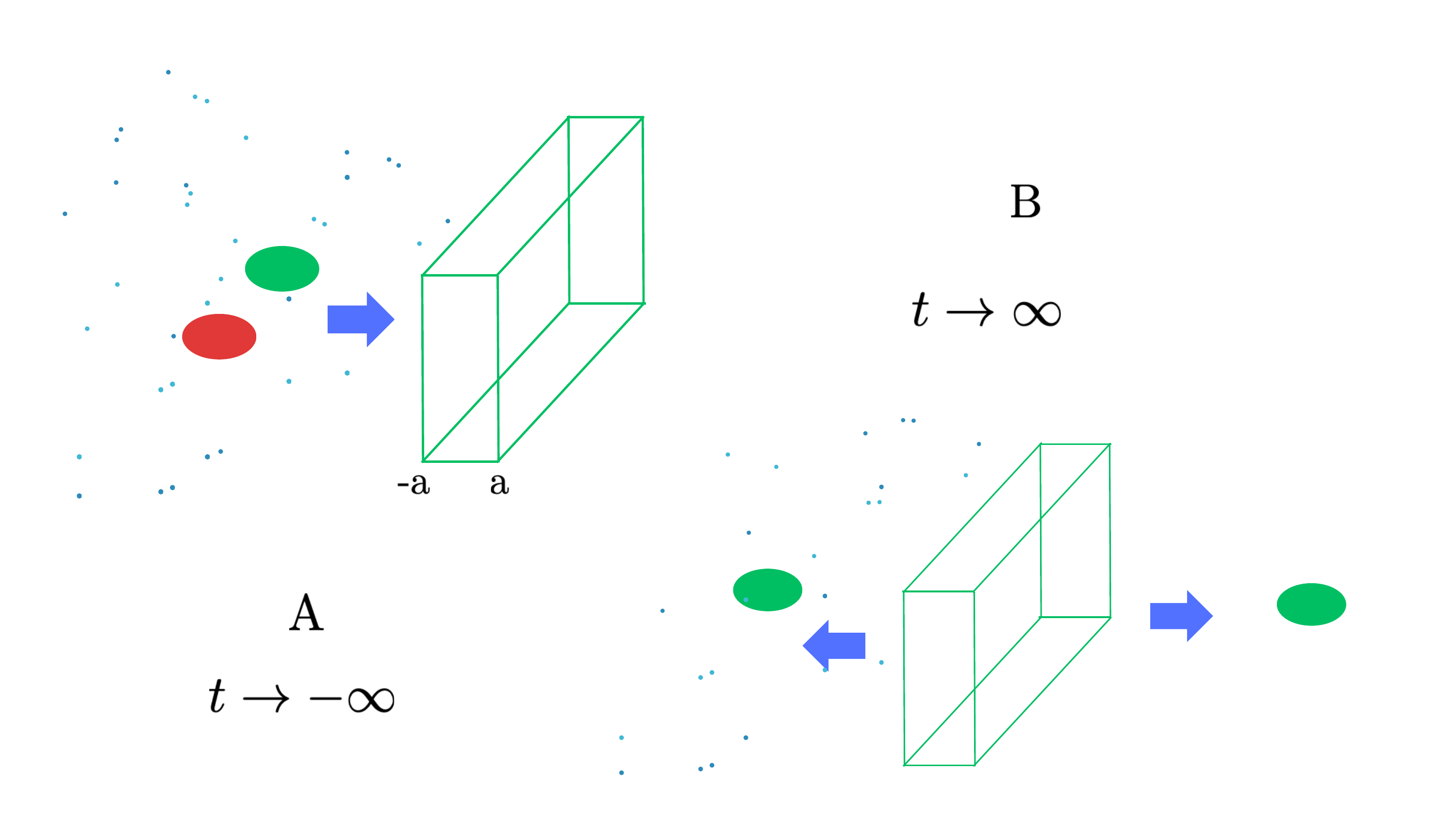}
\caption{\footnotesize  On the left figure (A), an atom (molecule) arrives from $z=-\infty$ and moves towards the potential barrier at $ -a < z < a$, as in  Fig.~\ref{ET1}.  But in contrast to the process in the vacuum in Fig.~\ref{ET1},  this time the half space on the left of the  potential barrier contains air.  The molecule is now in a mixed state due to the environment-induced decoherence. Its (transverse) position density matrix became diagonal: it is either near  $(x_1,y_1)$  or   near  $(x_2,y_2)$.
  The wave function  of the particle   at $t \to \infty$,  shown in the right part (B), contains still a small transmitted wave as well as the reflected wave, however without  
coherent superposition of two transverse wave packets.  }
\label{ET2}
\end{center}
\end{figure}

\section{Abstract concept of  ``particle of mass $m$"  \label{particle}}

It is customary to consider an otherwise unspecified  ``particle   of mass $m$", to discuss model systems, both in quantum mechanics and in classical mechanics. We will see that  the consideration based on such an abstract concept of a particle 
cannot be used to discriminate  classical objects from quantum mechanical systems, or as a way to explain the emergence of classical mechanics from QM.

Let us consider a $1D$ particle  of mass $m$,   moving in a harmonic-oscillator potential
\be   H =    \frac{p^2}{2m} +    \frac{ m \,\omega^2}{2}   x^2\;.
\ee
The coherent state
is defined by  
 $   a \,| \beta \ckt = \beta\,  |\beta \ckt\;,$  where $a$ is the annihilation operator. 
Its well-known solution,    in the coordinate representation, 
 is just the Gaussian wave packet  \footnote{ The position and momentum of the center of the wave packet  are  
\be 
x_0 =\sqrt{\dfrac{\hbar}{2 \, m\, \omega}}(\beta + \beta^{*}) \equiv
A\,\cos\varphi\;,  \quad 
 p_0  = i \sqrt{\dfrac{\hbar\, m \omega}{2}} ( \beta^{*}- \beta ) =
m\,\omega \, A \, \sin\varphi \;, \label{statocoer3} \ee
  }
\be \psi(x)=    \brc x |\beta\ckt = 
{\cal N} \exp \left[-\frac{(x-x_0)^2 }{4 \, D }  + i \, \frac{p_0  \, x}{\hbar}\right]\;,
\label{statocoerinSchr}
\ee
with  
\be   D= \brc (\Delta x)^2 \ckt =\frac{ \hbar}{2\,m\,\omega}\;.  
\label{coerente}  \ee
The  Schr\"odinger   time  evolution can be expressed as the time variation of the center of mass and its mean momentum,
$x_0 \to x_0(t)\;, \,\, p_0 \to p_0(t)\;, $
\be x_0(t) = A\, \cos(\varphi + \omega t)\qquad p_0(t) = m\,\omega\, A\,\sin(\varphi + \omega t)\;.
\label{statocoer5}\ee
while the wave packet shape and size (\ref{coerente})  remain  unchanged in time.   
This looks exactly like the motion of a classical oscillator of mass $m$ and  size $D$!

  It is sometimes thought that such a behavior of the coherent states carries the key to understand 
the emergence of classical mechanics  from QM.  There are however reasons to believe that this may not be quite the correct way of reasoning.

In order to see that such an identification/analogy  cannot be pushed too far,  consider  the quenching, i.e.,    turning off the oscillator potential  suddenly,   
setting  $\omega=0$, at $t=t_0$.   The particle starts moving freely, with the initial condition, 
$  (x_0(t_0), p_0(t_0))$. 

The problem is that there is no way to tell what happens at $t \ge t_0$. 
A quantum mechanical particle would diffuse, with the 
 rate depending  on mass, as in Table.~\ref{diffusion}.    
A classical particle does not diffuse. The expression  ``a particle of mass $m$"  does not 
tell if it is a quantum or a classical particle,  and what the true size of the body, $L_0$  (unrelated to $D$)   is.

 Note that by describing this body as a ``particle",  one has  tacitly assumed  that its physical size is irrelevant (i.e.,   $L_0=0$)   to the model harmonic-oscillator problem.  But the physical size of the particle, $L_0$, does matter.
  If  its (unspecified) size were truly zero,  it would be quantum mechanical, as $Q=R_q/L_0 = \infty$.

   The lesson one draws from this discussion is that 
a model system based on an abstract ``particle"  concept,   in which the information about $L_0$ is lacking,  
cannot be used  to study the  emergence of classical physics from quantum mechanics.  
Allowing for the decoherence effects, and selecting particular class of the mixed states as privileged ones  by introducing some criteria, 
may not lead to  a satisfactory understanding of how classical physics emerges from QM.

\section{Discussion  \label{Last}}

An immediate implication of the introduction of the concept of quantum ratio  is that the elementary particles are quantum mechanical.  This is so even if under certain conditions such as envirornment-induced decoherence, they may be reduced to mixed states. They remain quantum mechanical. 
The distinction between the concept of mixed (quantum) states and classical states is essential.  As the electron and photon are elementary particles 
 they remain quantum mechanical even in warm and dense environment of biological systems. 

We have studied the quantum ratio of some larger particles, atoms and molecules,  via examination of various interferometry experiments,  
which indeed show that these particles behave quantum-mechanically, in the vacuum. 

We discussed  extensively   in  Sec.~\ref{Decoherence}  several real and model examples involving atoms, molecules and elementary particles, to highlight the reasons 
why the environment-induced decoherence \cite{Joos1}-\cite{Zurek2},  in itself,   does not make the 
particle affected classical, as often stated or  assumed tacitly.  

Though in a slightly different context, the so-called negative-result experiments or  null measurements \cite{Renninger, BombTester}    tell us a 
 similar story.   
There,  the exclusion of some of the possible experimental outcomes (a non-measurement),  by use of an intentionally biased measurement set-up, implies the loss of the original  superposition of states. 
But the predicted state of the system is still perfectly a quantum-mechanical one,  though it now lives in a more restricted region of the Hilbert space.  See \cite{KK3} for a recent review  and for a careful re-examination  of the interpretation of  these negative-result experiments.

All these discussions lead us  naturally back  to the recurrent theme in quantum mechanics:  mixed state versus pure quantum states.  As is widely acknowledged, there are no  differences of principle.  As famously noted by Schr\"odinger,   a complete knowledge of the total, closed system  $\Sigma$    (its  wave function, the pure state vector) does not necessarily mean the same for   a part of the system    ($A \subset \Sigma$).

Only in exceptional situations in which the interactions {\it   and correlations}  between the subsystem of our interest (``local", $A$)  and the rest of the world  (``rest", $\Sigma/A$)  can be  neglected, and 
therefore  the total wave function has a
factorized form 
\be     \Psi^{\Sigma}   \simeq   \psi^{A}  \otimes   \Phi^{\Sigma/A}   \;,\label{factorization}  
\ee
 can we  describe  the local system $A$   in terms of  a  wave function.
Whenever the factorization (\ref{factorization}) fails,  system $A$  is  a mixture, described by a density matrix.

The quantum measurement is a process in which the factorized state (\ref{factorization}), where   $ \psi^{A}$ is the quantum state of interest,  and the measurement device $
\Phi^0$   is part of $\Phi^{\Sigma/A}$,   is brought into an entangled state,  triggered by a spacetime pointlike interaction event   \cite{KK1,KKTalk,KK3}.

Even a macroscopic system can be brought to the pure-state form $\psi^A$  as in  (\ref{factorization}), at sufficiently low temperatures.  At $T=0$  any system is  in its  quantum mechanical ground state.   See    \cite{Leggett}-\cite{Aaron}
  for the efforts to realize such macroscopic quantum states experimentally by going to very low temperatures. 

Vice versa,  classical equation of motion describes the CM of a macroscopic body {\it  at finite temperatures.}   When three conditions 
\begin{itemize}
  \item[(i)] that for macroscopic motions  (i.e., $\hbar \simeq 0$)  the Heisenberg relation does not limit the simultaneous determination - the initial condition - for the position  and momentum;
  \item[(ii)]   the lack of quantum diffusion, due to a large mass (a large number of atoms and molecules composing the body);  and
  \item[(iii)]  a finite body temperature, implying the thermal decoherence and mixed-state nature of the body,   
\end{itemize}
are met,  the CM of a body has a  unique trajectory  \cite{KK2}. 
Newton's equations for it  follow from the Ehrenfest theorem.  If the quantum fluctuation range $R_q$ 
is not larger than the size of the body, i.e.,  $Q=  R_q / L_0 \lesssim  1$, then such a trajectory can be regarded as the classical trajectory of that body.

To summarize, introduction of the quantum ratio is an attempt to go beyond the familiar ideas on the emergence of classical physics from QM, 
such as  the large action, semiclassical  limit (or $\hbar \to 0$)  and  Bohr's correspondence principle, or   
the environment-induced decoherence \cite{Joos1}-\cite{Zurek2}.
 Clearly, there is no 
sharp boundary between where or when  QM or Classical mechanics describes a given system more appropriately.  The quantum ratio is a proposal for an approximate, but  simple, universal criterion for characterizing the two kinds of physical systems:   quantum  ($Q \gg 1$)   and classical  ($Q \lesssim 1$).

\section*{Acknowledgments}   

  We thank   Francesco Cappuzzello, Giovanni Casini, Marco Matone, Pietro Menotti and Arkady Vainshtein  for discussions. 
  The work by KK is supported by the INFN special-initiative project  grant  GAST (Gauge and String Theories).

{}

\appendix


%
%
%
%
%
%
%
%
%
%
%
%

\section{Variational solution for the SG wavepackets  \label{Variational} } 

The Schr\"odinger equation  (\ref{SEq}) can be solved by separation of the variables,
\be  \Psi({\mathbf r}, t) =   \chi(x,t) \eta(y, t) \psi(z,t)  \;.
\ee 
As the motions in $x$ and $y$ directions are free ones,  we shall focus on $\psi(z,t).$

We recall  Dirac's variational principle \cite{Jackiw2}.
Consider the effective action
\be
\Gamma[\psi] = \int dt\;\langle\psi (t)\vert (i\partial_t - \hat{H})\vert
\psi (t)\rangle\;.  \label{action}  
\ee
 The variation with respect to  $|\psi\ckt$  and  $\brc \psi| $  
\begin{equation}
\label{g15}
\frac{\delta\Gamma[\psi ]}{\delta\psi} = 0\;\; ,\;\;\;\mbox{for all}\;\psi
\;\;\mbox{with}\;\langle\psi\vert\psi\rangle = 1\;\; ,
\end{equation}
i.e., requiring the effective action $\Gamma$ to be stationary against 
arbitrary variations of the normalized wave
function, which vanish at $t \rightarrow \pm \infty$,    is equivalent to
the exact Schr\"odinger equation \footnote{This has been applied  in a study of semiquantum chaos 
in a double-well oscillator in Ref.\,\cite{BlumElze}, but can be used  as well 
in quantum field theory with suitable wave functionals:     see  \cite{Jackiw2,Elze}, for example.}.

An important property, which follows immediately, is that orthogonal superpositions of variational trial eigenstates of the Hamiltonian evolve independently, without interfering with each other. Let 
$|\psi\rangle =|\psi_1\rangle |\!\uparrow\rangle 
+|\psi_2\rangle |\!\downarrow\rangle\rangle$  be the sum of the two orthogonal spin-up and spin-down eigenstates of
the simplified Hamiltonian \cite{Platt} 
\begin{equation}\label{SGHamiltonian} 
\hat{H}=\frac{1}{2m}\hat p_z^{\; 2}+\mu \, b_0\,  z  \,  \sigma_z\;, \qquad   \mu \equiv  \mu_B/2
\;\;   \end{equation} 
in the  Stern-Gerlach set-up   (with magnetic field of the type, (\ref{magneticfield})).
Then, the effective action becomes a sum of two independent terms, 
$\int \langle\psi_1 |\dots |\psi_1 \rangle$ and 
$\int\langle\psi_2 |\dots |\psi_2 \rangle$, which can be varied separately.  

We choose the following normalized  Gaussian trial wave functions, which are suitable for the effectively one-dimensional problem 
of particles with mass $m$ and the magnetic moment $\mu$ moving in a magnetic field  with gradient  transverse to the beam direction  (i.e., $\hat x$):    
\begin{equation}     
\psi (z,  t)=(2\pi G(t))^{-\frac{1}{4}}\exp\Big \{
-  \left( \tfrac{1}{4 G(t)}-i\sigma (t)\right)  \left(z-\bar z(t)\right)^2
		     +i\bar p(t) \left(z-\bar z(t)\right) \Big \}    \label{variationalwp}
 \end{equation} 
where  $G(t), \sigma(t), {\bar p}(t), {\bar z}(t)$  are the variational-parametric functions.  ${\bar p}(t), {\bar z}(t)$ describe the momentum and position of the wave packet, and  $G(t), \sigma(t)$
the quantum diffusion. 
Substituting this into  (\ref{action})  yields  
\begin{equation}\label{g18} 
\Gamma[\psi] = \int dt\left\{\bar p\dot{\bar z}-
\frac{1}{2m}\bar p^2 \mp\mu b_0  \bar z  
+\hbar \Big [\sigma\dot{G}-\frac{2}{m}\sigma^2G-\frac{1}{8m}G^{-1}\Big ]\right\}
\;.\end{equation}

Independent variations with respect to  $G(t), \sigma(t), {\bar p}(t), {\bar z}(t)$  give  
\begin{eqnarray}
\dot{\bar z}&=&\frac{1}{m}\bar p \;\;,
\qquad   
\dot{\bar p}=\mp  \mu  b_0 \;\;,
\label{g22} \\
\dot{G}&=&\frac{4}{m}\sigma G \;\;,
\qquad   
\dot{\sigma}=-\frac{2}{m}\sigma^2+\frac{1}{8m}G^{-2} \;\;. 
\label{g24}
\end{eqnarray}
Note that in the magnetic field of linear inhomogenuity   $ {\mathbf B}_z =  b_0\, z$  under consideration, the center of the wave packet    ${\bar p}(t), {\bar z}(t)$  moves as a classical particle, and  
the diffusion effects  $G(t), \sigma(t)$  are the same as  for a free wave packet.  

The solution of Eqs.\,(\ref{g22})-(\ref{g24}) is: 
\begin{eqnarray}
\bar z(t)  &=&\frac{1}{m}(\mp\frac{1}{2}   \mu \, b_0 \,  t^2+\bar p_0t)+\bar z_0 \;;  \qquad   
\bar p(t) = \mp \mu\, b_0 \, t+\bar p_0 \;\;,
\nonumber   \\
G(t) &=&\frac{i}{m}t+G_0 \;\;,  \qquad  
\sigma(t) =\frac{i}{4}(\frac{i}{m}t+G_0)^{-1} \;\;,  
\label{g34}
\end{eqnarray} 
where $\bar x_0$, $\bar p_0$ and  $G_0$     set the initial conditions at $t=0$. 
The diffusion of the wave packet,   
$\frac{1}{4}G^{-1}-i\sigma =\frac{1}{2}(\frac{i}{m}t+G_0)^{-1}$,   is  the same as in the free case as we noted already.
We understand this as due to the fact that in the linear field  $B_z(z) =  b_0\, z$
the force is constant and  the same for each part inside the wave packets,  $\psi_{1,2}(z)$.    The effect of quantum diffusion is negligible  for  $t \ll    m  G_0$,   where $G_0$
is the initial wave packet size. 

Substituting (\ref{g34})  into (\ref{variationalwp}) yields our variational solution  of the Schr\"odinger equation.
The wave packets 
for spin-up and spin-down states remain  in coherent superposition but move  independently   (see Fig.~\ref{WSG}).

\end{document}